\documentstyle[12pt,psfig]{article}
\input xy
\xyoption{all}

\begin{document}

\newcommand\M[2]{{\overline M}_{{#1},{#2}}}
\newcommand\th{\scriptstyle{{\rm th}}}

\hfill hep-th/0605005

\vspace{1.0in}

\begin{center}

{\large\bf Notes on Certain Other (0,2) Correlation Functions }

\vspace{0.75in}

Eric Sharpe \\
Departments of Physics, Mathematics \\
University of Utah \\
Salt Lake City, UT  84112 \\
{\tt ersharpe@math.utah.edu} \\

$\,$

\end{center}

In this paper we shall describe some correlation function computations
in perturbative heterotic strings that generalize B model computations.
On the $(2,2)$ locus, correlation functions in the B model receive
no quantum corrections, but off the $(2,2)$ locus, that can change.
Classically, the $(0,2)$ analogue of the B model is equivalent to
the previously-discussed $(0,2)$ analogue of the A model, but with the
gauge bundle dualized -- our generalization of the A model,
also simultaneously generalizes the B model.
The A and B analogues sometimes have different regularizations, however, which 
distinguish them quantum-mechanically.
We discuss how properties of the $(2,2)$ B model, such as the
lack of quantum corrections, are realized in $(0,2)$ A model language.
In an appendix, we also extensively discuss how the Calabi-Yau condition
for the closed string B model (uncoupled to topological gravity) can
be weakened slightly, a detail which does not seem to have been
covered in the literature previously.  That weakening also manifests in the
description of the $(2,2)$ B model as a $(0,2)$ A model.

\begin{flushleft}
April 2006
\end{flushleft}

\newpage

\tableofcontents

\newpage

\section{Introduction}

Understanding rational curve corrections in heterotic compactifications
in which the gauge bundle is different from the tangent bundle has been
a technical problem for many years.  This problem resurfaced a few years
ago in new attempts \cite{abs} to understand (0,2) mirror symmetry.
Partly in order to check some conjectures made in \cite{abs} regarding
analogues of $\overline{\bf 27}^3$ couplings, the (0,2) analogue of
the A model was worked out in \cite{ks}.
(See \cite{ksa} for a review.)

Although a nonlinear sigma model with (0,2) supersymmetry cannot be
twisted to a true topological field theory, one can perform an analogous
twist, which reduces to the ordinary A model on the (2,2) locus, and as
discussed in \cite{ks}, the resulting physical theory has many of the
same general features as the ordinary A model.  

Mathematically, the work in \cite{ks} defines something analogous to,
but distinct from, Gromov-Witten invariants.  Instead of intersection
theory on a moduli space of curves, one instead calculates sheaf
cohomology on such a moduli space, and so the results of calculations
in the (0,2) analogues can be understood as quantum-corrected sheaf
cohomology.

The work \cite{ks} also helped establish the existence of a heterotic
version of a quantum cohomology ring.  The work \cite{abs} made a conjecture
for the form of such a ring structure; our work \cite{ks} provided
concrete calculations of correlation functions that supported the existence
of such rings; and then recently \cite{ad} provided a general argument
explaining why such rings should exist.
More recently \cite{tan1} studied perturbative aspects of the (0,2)
A models from the point of view of chiral differential operators.

Having studied the (0,2) analogue of the A model extensively in
\cite{ks}, in this paper we examine the (0,2) analogue of the B model,
and analogues of ${\bf 27}^3$ couplings.  Although the B model does not
receive quantum corrections on the (2,2) locus, by contrast off the
(0,2) locus the (0,2) analogue of the B model can and does receive
quantum corrections.

Along the way, we shall learn some important things about the ordinary
B model.  For example, why is the ordinary B model classical?
For {\it most} rational curves, an index theory argument shows that the
number of fermi zero modes does not match correlators, and hence
no quantum corrections, but there are some exceptional cases in which
such arguments do not apply.

Furthermore, there is a (classical) symmetry between the (0,2) A and B models:
by exchanging the gauge bundle ${\cal E}$ with its dual ${\cal E}^{\vee}$,
one exchanges the (0,2) A and B model twists.
Now, there exist rational curves for which $\phi^* TX \cong \phi^* T^*X$,
and for those curves, one would therefore expect that their contributions
to the A and B model would be identical -- yet the A model gets nonvanishing
corrections in such sectors, while the B model is classical.
In fact, these exceptional cases, are the same ones mentioned above -- 
index theory does not suffice in these examples to show that their
contributions to the B model vanish.

In such sectors, we find that two possible things happen.
In one set of cases,
because the isomorphisms 
 $\phi^* TX \cong \phi^* T^*X$ fit together nontrivially over the moduli space,
the A and B models are actually distinct classically,
although the contribution from single rational curves appears to
be the same -- our analysis above was a little too naive.

In another set of cases, quantum corrections to both the
A and B model are identical classically, on the part of the moduli space
corresponding to honest maps, and have the form
\begin{displaymath}
\int_{ {\cal M} } \alpha
\end{displaymath}
where $\alpha$ is a top-form on ${\cal M}$.
However, a purely classical analysis does not suffice to define the theory,
one must regularize it -- which in this case means compactifying the moduli
space and extending sheaves over the compactification divisor.
It turns out there are two distinct regularizations of the A and B model
in these exceptional cases.  The regularization natural to the A model
yields a nonvanishing result, whereas the regularization natural to the
B model yields a vanishing result.

In addition to learning at a deeper level why the B model is classical,
we shall also learn that the B model can be defined on slightly
more spaces than just Calabi-Yau's:  the B model can also be defined on
spaces on which $K^{\otimes 2} \cong {\cal O}$.

We begin in section~\ref{outlineB} by outlining how to perform
calculations in the (0,2) B model.  The details are extremely similar
in spirit to those performed for the (0,2) A model in \cite{ks}.
In section~\ref{Bon22} we discuss how calculations for the (0,2) B model
specialize on the (2,2) locus to the purely classical results of
the ordinary B model.  In section~\ref{relnAB} we discuss the relation
between the (0,2) analogues of the A and B models.
Classically the two twists can be exchanged by replacing the gauge
bundle ${\cal E}$ with its dual ${\cal E}^{\vee}$, but a purely classical
analysis does not always suffice:  to fully understand the story one must also
understand that there are sometimes two regularizations of the quantum field 
theory,
two different consistent ways to extend induced sheaves over 
the compactification
divisor, which distinguish the A and B analogue twists.
In section~\ref{nonrenorm} we discuss suggestions of a nonrenormalization
result for these calculations.  The consistency conditions for the
(0,2) B model imply that on the (2,2) locus, one need merely require
only that $K^{\otimes} \cong {\cal O}$, a different result that is usually
stated.  In appendix~\ref{bmod} we discuss how this is consistent,
and in fact implicit in old B model calcuations.
Finally, in section~\ref{anomanalysis} we comment on how the anomaly
analysis used here applies to the A and B models in instanton sectors at
arbitrary genus.

We should also briefly mention some recent work on correlation functions
of gauge singlets in heterotic compactifications.
The very interesting paper \cite{beased} 
argued that a certain class of correlation functions
did not receive quantum corrections.  The correlation functions in question
all involve gauge singlets, whereas the correlation functions considered
in this paper and \cite{ks} are not gauge singlets.  
Whether a given correlation function receives quantum corrections or
not is a function of the correlators -- this is the reason why
physical amplitudes corresponding to $(2,2)$ A and B model amplitudes
do and do not receive quantum corrections, respectively.
There are some other significant differences between these works.
In \cite{beased}, as they were considering gauge singlet correlators,
they only performed a topological twist on worldsheet right-movers,
whereas we perform a (pseudo-)topological twist on both right- and
left-movers.  As a result, their correlation functions amount to
integrals of sections of the canonical bundle -- residues, in short,
in the sense of \cite{hart}[pp 247-248] and \cite{gh}[p. 731]
-- whereas in this paper and \cite{ks} correlation functions are
integrals of scalars.

\section{Outline of the $(0,2)$ B model computations}
\label{outlineB}

First, let us briefly recall the $(0,2)$ A model analogue calculation
from \cite{ks}.
Given a $(0,2)$ nonlinear sigma model on a complex K\"ahler manifold $X$
with a holomorphic vector bundle ${\cal E}$, obeying the constraints
\begin{eqnarray}
\Lambda^{top} {\cal E}^{\vee} & \cong & K_X    \label{constr1}\\
\mbox{ch}_2( {\cal E}) & = & \mbox{ch}_2(TX)   \label{constr2}
\end{eqnarray}
the A analogue twist is defined by coupling the worldsheet fermions to
the following bundles:
\begin{eqnarray*}
\psi_+^i & \in & \Gamma_{ C^{\infty} }\left( \phi^* T^{1,0}X \right) \\
\psi_+^{\overline{\imath}} & \in &
\Gamma_{ C^{\infty} }\left( K_{\Sigma} \otimes \left(
\phi^* T^{1,0}X \right)^{\vee} \right) \\
\lambda_-^a & \in & \Gamma_{ C^{\infty} }\left(
\overline{K}_{\Sigma} \otimes \left( \phi^* \overline{\epsilon} \right)^{\vee} \right) \\
\lambda_-^{\overline{a}} & \in &
\Gamma_{ C^{\infty} }\left( \phi^* \overline{\epsilon} \right)
\end{eqnarray*}

The B analogue twist is defined by making the opposite choice for the
left-moving fermions:
\begin{eqnarray*}
\psi_+^i & \in & \Gamma_{ C^{\infty} }\left( \phi^* T^{1,0}X \right) \\
\psi_+^{\overline{\imath}} & \in &
\Gamma_{ C^{\infty} }\left( K_{\Sigma} \otimes \left(
\phi^* T^{1,0}X \right)^{\vee} \right) \\
\lambda_-^a & \in & \Gamma_{ C^{\infty} }\left(
\left( \phi^* \overline{\epsilon} \right)^{\vee} \right) \\
\lambda_-^{\overline{a}} & \in &
\Gamma_{ C^{\infty} }\left( \overline{K}_{\Sigma} \otimes
\phi^* \overline{\epsilon} \right)
\end{eqnarray*}
(The observant reader will note that although our conventions match those
of \cite{edtft} for the A model, our conventions are the opposite of his
for the B model.  Unfortunately, to make the analysis of duality
symmetries manifest, such a convention mismatch is unavoidable.)

In the A analogue twist we defined a sheaf ${\cal F}$ of the 
$\lambda_-^{\overline{a}}$ zero modes, given in a large open patch on the
moduli space ${\cal M}$ by 
\begin{displaymath}
{\cal F} \: = \: R^0 \pi_* \alpha^* {\cal E}
\end{displaymath}
(corresponding to the fact that for any given worldsheet instanton,
the zero modes are counted by $H^0(\Sigma, \phi^* {\cal E})$)
where $\alpha: \Sigma \times {\cal M} \rightarrow X$ is the universal
map and $\pi: \Sigma \times {\cal M} \rightarrow {\cal M}$ is the
projection.
The classical correlators, elements of $H^{\cdot}(X, \Lambda^{\cdot} 
{\cal E}^{\vee})$, define elements of $H^{\cdot}({\cal M}, \Lambda^{\cdot}
{\cal M}^{\vee})$ quantum-mechanically.
The sheaf of the other left-moving fermion zero modes is denoted
${\cal F}_1$ and is given, on a large open patch on ${\cal M}$, by
\begin{displaymath}
{\cal F}_1 \: = \: R^1 \pi_* \alpha^* {\cal E}
\end{displaymath}
(corresponding to the fact that for any given worldsheet instanton,
the zero modes are counted by $H^0(\Sigma, K_{\sigma} \otimes
\phi^* {\cal E}^{\vee}) \cong H^1( \Sigma, \phi^* {\cal E} )$).
The sheaf ${\cal F}_1$ plays a role closely analogous to the obstruction
bundle in standard A model computations, as described in detail in
\cite{ks}.

On the $(2,2)$ locus, ${\cal E} = TX$, the sheaf ${\cal F}$ becomes
$T {\cal M}$, and ${\cal F}_1$ becomes the obstruction sheaf,
henceforward denoted Obs.  The classical correlators are given
by $H^{\cdot}(X, \Lambda^{\cdot} T^* X) = H^{\cdot, \cdot}(X)$,
and similarly for the quantum correlators.

We can define the $(0,2)$ B analogue correlation functions similarly.
In the B analogue twist, the role of the sheaf ${\cal F}$ is replaced by
\begin{displaymath}
{\cal F}_B \: = \: R^0 \pi_* \alpha^* \left( {\cal E}^{\vee} \right)
\end{displaymath}
as ${\cal F}$ in either incarnation is the sheaf over ${\cal M}$ of
zero modes of the left-moving scalars, which for the B analogue
twist are the 
\begin{displaymath}
\lambda_-^a \: \in \: \Gamma_{ C^{\infty} }\left( \phi^*
\overline{ {\cal E} }^{\vee} \right)
\end{displaymath}
Similarly, the role of ${\cal F}_1$ is replaced by
\begin{displaymath}
{\cal F}_{B 1}  \: = \: R^1 \pi_* \alpha^* \left( {\cal E}^{\vee} \right)
\end{displaymath}
as ${\cal F}_1$ in either incarnation is the sheaf over
${\cal M}$ of zero modes of the left-moving vectors, which for the
B analogue twist are the
\begin{displaymath}
\lambda_-^{\overline{a}} \: \in \:
\Gamma_{ C^{\infty} }\left( \overline{K}_{\Sigma} \otimes
\left( \phi^* \overline{ {\cal E} }\right)^{\vee} \right)
\end{displaymath}
The classical correlators in the B analogue twist correspond to 
elements of $H^{\cdot}(X, \Lambda^{\cdot} {\cal E} )$
(which on the $(2,2)$ locus become elements of
$H^{\cdot}(X, \Lambda^{\cdot} TX)$, and so the quantum correlators
must be elements of $H^{\cdot}({\cal M}, \Lambda^{\cdot} {\cal F}_B^{\vee})$,
as this correctly reproduces the classical limit.
Correlation functions in the B model analogue in the case that both
the obstruction sheaf and ${\cal F}_{B 1}$ vanish are computed in
the form
\begin{displaymath}
< {\cal O}_1 \cdots {\cal O}_r > \: = \:
\int_{ {\cal M} } H^{top}\left( {\cal M}, \Lambda^{top} {\cal F}_B^{\vee} \right)
\end{displaymath}
just as in the A analogue twist \cite{ks}, and the effects of four-fermi
terms brought in to soak up the excess zero modes represented by
the obstruction sheaf and ${\cal F}_{B 1}$ can be reproduced in the obvious 
form
closely analogous to that described in \cite{ks}.
Just as in \cite{ks}, the fact that one is integrating a top-form over
${\cal M}$ is ultimately a consequence of the anomaly-cancellation condition,
together with the constraint that the path-integral-measure in the twisted
theory is nonanomalous.

We have been brief in our discussion of the computation of
correlation functions in the $(0,2)$ B model analogue, because
there is a more efficient\footnote{We would like to thank
J.~Distler for pointing this out to us.} way to think about them.
Note that the difference between the A and B analogue twists of the
$(0,2)$ theory can be reversed by replacing ${\cal E}$ with
${\cal E}^{\vee}$.  This has the effect of exchanging
$\lambda_-^a$ and $\lambda_-^{\overline{a}}$, which in our conventions
precisely exchanges the A and B twists. 
Hence, a B analogue correlation function for a space $X$
with bundle ${\cal E}$ can be computed as an A analogue correlation
function for $X$ with bundle ${\cal E}^{\vee}$, 
something that can be independently checked by comparing
the details of the B analogue computations above to the
A analogue computations discussed in \cite{ks}.
In effect, in our previous work \cite{ks} on the A model analogue,
we already also computed the B analogue correlation functions.

Strictly speaking, this conclusion is true of the A, B twists classically.
Later we shall see that if this were also true quantum-mechanically,
one could sometimes formulate contradictions involving curves $\phi$ such that
$\phi^* T \cong \phi^* T^*$.  In some such cases, classically 
the A and B models get the same contributions, yet there are examples
in which A model contributions are nonzero while the B model remains
classical.  The resolution of this puzzle lies in the regularizations of
the quantum field theories:  in such examples, there are two distinct
regularizations of the physical two-dimensional quantum field theory,
corresponding to distinct extensions of induced sheaves over the
compactified moduli space ${\cal M}$, and the different regularizations are
responsible for getting different results.

In any event, for the most part the techniques of our previous 
work \cite{ks} are
immediately applicable here.  For example, in \cite{ks} we extensively 
discuss the issue of compactification of moduli spaces and how to go
about extending induced sheaves over compactification divisors so as
to get results possessing needed symmetries.  We have not reviewed that
material here, but as the analysis is identical to that in
\cite{ks}, will assume it henceforward.

\section{The B model analogue on the $(2,2)$ locus}
\label{Bon22}

From our previous discussion, the $(2,2)$ B model on $X$ should be
equivalent to the $(0,2)$ A model analogue on $X$ with
gauge bundle given by $T^* X$.  Let us perform some checks
of that statement.

First, note that the consistency conditions
\ref{constr1} and \ref{constr2} reduce to the anomaly-cancellation
condition for the closed string B model (see appendix~\ref{bmod}).
Specifically, when ${\cal E} = T^* X$, the constraint~(\ref{constr1})
becomes the condition
\begin{displaymath}
\Lambda^{top} TX \: \cong \: K_X
\end{displaymath}
or, equivalently, $K_X^{\otimes 2} \cong {\cal O}_X$.
Ordinarily one says that the B model is only well-defined on
Calabi-Yau's, {\it i.e.} spaces for which $K \cong {\cal O}$,
but in fact we argue in appendix~\ref{bmod} that for the closed
string B model to be anomaly-free merely requires the weaker condition
above.  As we discuss the matter extensively in the appendix,
here we shall move on.

The second constraint~(\ref{constr2}) is trivially satisfied under
these circumstances.  Thus, the A model analogue consistency conditions
become the B model consistency conditions when 
the A model analogue is computed with ${\cal E} = T^* X$,
exactly as one would hope.

Another important property of the $(2,2)$ B model is that it
has no quantum corrections.  In the present language, that fact
is much more obscure.  
For ``most'' cases, this can be established by index theory arguments,
but there are some special cases in which index theory arguments fail.
In the remainder of this section, we shall investigate this matter,
first by reviewing the index theory arguments and places where they
break down, then considering some specific examples of breakdowns.
What we shall find is that although classically it looks as if there
could be quantum corrections in such sectors, when we regularize the theory
(by compactifying moduli spaces and extending sheaves over compactification
divisors), the potential quantum corrections take the form of the
integral of an exact form, and so vanish.

We shall begin with a general discussion showing how index theory
can be used to understand many, but not all, cases, then consider
some examples of special cases not covered by index theory.
Additional examples not covered by index theory can be found in
\cite{twoparam1,kmp}, concerning hypersurfaces in
${\bf P}_{1,1,2,2,2}$.

\subsection{General vanishing results}

We shall begin by considering how index theory can be used to
rule out quantum corrections in most (though not all) cases.
Let us break up the problem into cases, according to the structure
of the normal bundle.
\begin{enumerate}
\item Suppose 
\begin{displaymath}
TX|_{{\bf P}^1} \: = \: {\cal O}(2) \oplus {\cal O}(-1) \oplus {\cal O}(-1)
\end{displaymath}
as appropriate for an isolated rational curve.
Then
\begin{eqnarray*}
\phi^* T & = & {\cal O}(2d) \oplus {\cal O}(-d) \oplus {\cal O}(-d) \\
\phi^* T^* & = & {\cal O}(-2d) \oplus {\cal O}(d) \oplus {\cal O}(d)
\end{eqnarray*}
hence
\begin{eqnarray*}
\mbox{rank } \mbox{Obs} & = & h^1\left( {\bf P}^1, \phi^* T \right) \: = \:
2(d - 1) \\
\mbox{rank } {\cal F}_1 & = & h^1\left( {\bf P}^1, \phi^* T^* \right) \: = \:
2d - 1
\end{eqnarray*}
These ranks do not match, meaning that the number of excess $\psi_+$ zero
modes does not match the number of excess $\lambda_-$ zero modes,
and so in particular four-fermi terms alone cannot resolve the discrepancy.
There is no contribution to correlation functions from purely zero
modes in this sector, and supersymmetry prevents other corrections,
hence in this sector correlation functions must vanish.
\item Suppose
\begin{displaymath}
TX|_{ {\bf P}^1} \: = \: {\cal O}(2) \oplus {\cal O}(1) \oplus {\cal O}(-3)
\end{displaymath}
Then
\begin{eqnarray*}
\phi^* T & = & {\cal O}(2d) \oplus {\cal O}(d) \oplus {\cal O}(-3d) \\
\phi^* T^* & = & {\cal O}(-2d) \oplus {\cal O}(-d) \oplus {\cal O}(3d)
\end{eqnarray*}
hence
\begin{eqnarray*}
\mbox{rank } \mbox{Obs} & = & h^1\left( {\bf P}^1, \phi^* T \right) \: = \:
3d -1 \\
\mbox{rank } {\cal F}_1 & = & h^1\left( {\bf P}^1, \phi^* T^* \right) \: = \:
(2d - 1) \: + \: (d - 1 )
\end{eqnarray*}
Again, there is a mismatch between left- and right-moving excess zero modes,
which cannot be cured with four-fermi terms.  As in the previous case,
this sector cannot contribute to correlation functions.
\item Suppose
\begin{displaymath}
TX|_{ {\bf P}^1 } \: = \: {\cal O}(2) \oplus {\cal O} \oplus {\cal O}(-2)
\end{displaymath}
(This is essentially the case we saw in the previous analysis of
$\widetilde{ {\bf C}^2/{\bf Z}_2 }$, modulo the difference in dimension.)
Then
\begin{eqnarray*}
\phi^* T & = & {\cal O}(2d) \oplus {\cal O} \oplus {\cal O}(-2d) \\
\phi^* T^* & = & {\cal O}(-2d) \oplus {\cal O} \oplus {\cal O}(2d) \\
 & \cong & \phi^* T
\end{eqnarray*}
hence
\begin{eqnarray*}
\mbox{rank }\mbox{Obs} & = & h^1\left( {\bf P}^1, \phi^* T \right) \: = \:
2d - 1 \\
\mbox{rank } {\cal F}_1 & = & h^1\left( {\bf P}^1, \phi^* T^* \right) \: = \:
2d - 1 \\
& = & \mbox{rank }\mbox{Obs} 
\end{eqnarray*}
In this case, the number of left- and right-moving excess fermi zero modes
{\it does} match, unlike the last two cases, so in principle one
could expect to get a nonzero contribution by pulling down four-fermi
terms.  
\end{enumerate}
More generally, only for rational curves of this last form
(namely, $\phi^* T \cong \phi^* T^*$)
is it possible to have a nonzero contribution to a B-model-type
correlation function. 
In the next few sections we shall study several examples of such
cases.  We shall find that in all 
such examples appearing in this paper,
the four-fermi terms
contribute a cohomologically-trivial factor, forcing quantum
corrections to vanish.

\subsection{Example:  $\widetilde{ {\bf C}^2/{\bf Z}_2 }$ }   \label{c2z2B}

Consider the toric Calabi-Yau $\widetilde{ {\bf C}^2/ {\bf Z}_2 }$,
described as a gauged linear sigma model with chiral superfields
$x$, $y$, $p$ of charges $1$, $1$, $-2$, respectively, under a single
gauged $U(1)$.  It is easy to check that this gauged linear sigma
model describes a one-parameter family of resolution of 
${\bf C}^2/{\bf Z}_2$, with the size of the exceptional divisor determined
by the Fayet-Iliopoulos parameter.

Now, it turns out that both A and B models are classical on this space,
because it is a K3:  the A model is independent of complex structure, and
for generic complex structure, K3 surfaces are not algebraic, and so have
no rational curves.  For algebraic K3's, at a computational level, the
vanishing manifests itself as the fact that 
a degree-two cohomology class on the K3 will induce
a vanishing cohomology class on the moduli space, hence there can
be no quantum corrections.  Nevertheless, other computations
on algebraic K3 surfaces proceed as usual, and form a good simple
example of the general statement.  Thus, we shall work through the computation
formally, as the resulting moduli space and behavior of four-fermi terms
is a good prototype for other cases.

Although $\widetilde{ {\bf C}^2/{\bf Z}_2}$ itself is noncompact, 
the space of curves in the space
is often compact.
Recall the linear sigma model moduli space ${\cal M}$ is defined
by expanding each chiral superfield in a basis of zero modes,
and then interpreting the coefficient of each zero mode as a homogeneous
coordinate on the moduli space, of the same weight as the original
chiral superfield.
In the present case,
\begin{eqnarray*}
x, y & \mapsto & H^0\left({\bf P}^1, {\cal O}(d) \right) \: = \: {\bf C}^{d+1}
\mbox{ for } d > 0 \\
p & \mapsto & H^0\left( {\bf P}^1, {\cal O}(-2d) \right) \: = \: 0
\mbox{ for } d > 0
\end{eqnarray*}
We shall assume $d > 0$ for simplicity, so the moduli space ${\cal M}$
is given by ${\bf P}^{2(d+1)-1}$.

Next, we need to compute ${\cal F}$ and ${\cal F}_1$ over the moduli space.
Thinking of this as an A model analogue computation with gauge
bundle $T^* X$, we first describe the tangent bundle by
\begin{displaymath}
0 \: \longrightarrow \: {\cal O} \:
\stackrel{ [x,y,p] }{\longrightarrow} \:
{\cal O}(1)^{\oplus 2} \oplus {\cal O}(-2) \: \longrightarrow \: TX \:
\longrightarrow \: 0
\end{displaymath}
which is dualized to
\begin{displaymath}
0 \: \longrightarrow \:
T^* X \: \longrightarrow \:
{\cal O}(-1)^{\oplus 2} \oplus {\cal O}(2) \: 
\stackrel{ [x,y,p]^T }{\longrightarrow} \: {\cal O} \: \longrightarrow \: 0
\end{displaymath}
Following the prescription of \cite{ks}, we define ${\cal F}$ and ${\cal F}_1$
over ${\cal M}$ by the following long exact sequence:
\begin{displaymath}
\begin{array}{ccccccc}
0 & \longrightarrow & {\cal F} & \longrightarrow &
\left[ H^0\left( {\bf P}^1, {\cal O}(-d) \right) 
\otimes_{ {\bf C} } {\cal O}(-1) \right]^{\oplus 2} \oplus
H^0\left( {\bf P}^1, {\cal O}(2d) \right) \otimes_{ {\bf C} } {\cal O}(2)
& &  
\\
  & & & &
\stackrel{*}{\longrightarrow} \: H^0\left( {\bf P}^1, {\cal O} \right) 
\otimes_{ {\bf C} } {\cal O} & &   \\
 & \longrightarrow &
{\cal F}_1 & \longrightarrow &
\left[ H^1\left( {\bf P}^1, {\cal O}(-d) \right)
\otimes_{ {\bf C} } {\cal O}(-1) \right]^{\oplus 2} \oplus
H^1\left( {\bf P}^1, {\cal O}(2d) \right) \otimes_{ {\bf C} } {\cal O}(2)
& &  \\
 & & & &
\longrightarrow \: H^1\left( {\bf P}^1, {\cal O} \right) \otimes_{ {\bf C} }
{\cal O} \: \longrightarrow \: 0 & &
\end{array}
\end{displaymath}
For $d > 0$, the terms in the long exact sequence above simplify considerably.
A further simplification can be obtained by studying the map denoted
$*$.  This map is defined by expanding $x$, $y$, and $p$ in their
zero modes, then $[x,y,p]^T$ induces $*$.  However, for $d>0$,
there are no $p$ zero modes, so all elements of
\begin{displaymath}
H^0\left( {\bf P}^1, {\cal O}(2d) \right) \otimes_{ {\bf C} }
{\cal O}(2)
\end{displaymath}
are mapped to zero, and the $x$ and $y$ zero modes both act on copies
of
\begin{displaymath}
H^0\left( {\bf P}^1, {\cal O}(-d) \right) \otimes_{ {\bf C} }
{\cal O}(-1) \: = \: 0
\end{displaymath}
Thus, the map $*$ is zero.
As a result, we can write
\begin{equation}   \label{ff1}
\begin{array}{c}
{\cal F} \: \cong \: H^0\left( {\bf P}^1, {\cal O}(2d) \right) \otimes_{ {\bf C}}
{\cal O}(2) \\
0 \: \longrightarrow \: {\cal O} \: \longrightarrow \:
{\cal F}_1 \: \longrightarrow \:
\left[ H^1\left( {\bf P}^1, {\cal O}(-d) \right) \otimes_{ {\bf C} }
{\cal O}(-1) \right]^{\oplus 2} \: \longrightarrow \: 0
\end{array}
\end{equation}
For purposes of comparison, the obstruction sheaf in this case is given
by
\begin{displaymath}
\mbox{Obs} \: \cong \:
H^1\left( {\bf P}^1, {\cal O}(-2d) \right) \otimes_{ {\bf C} } {\cal O}(-2)
\end{displaymath}
as discussed in \cite{ks}.
It is straightforward to check that
\begin{eqnarray*}
\mbox{rank } {\cal F} \: = & 2d \: + \: 1 & = \: \mbox{rank } T {\cal M} \\
\mbox{rank } {\cal F}_1 \: = & 2d \: - \: 1 & = \: \mbox{rank }\mbox{Obs} \\
c_1\left( {\cal F} \ominus {\cal F}_1 \right) \: = & 6d J & = \:
c_1\left( T {\cal M} \ominus \mbox{Obs} \right)
\end{eqnarray*}
for $J$ the generator of $H^2( {\cal M}, {\bf Z} )$.

Let us now further restrict to the special case $d=1$.
Here,
\begin{eqnarray*}
{\cal M} & = & {\bf P}^3 \\
\mbox{Obs} & \cong & {\cal O}(-2) \\
{\cal F} & \cong & {\cal O}(2)^{\oplus 3} \\
{\cal F}_1 & \cong & {\cal O} 
\end{eqnarray*}
A potentially nonzero correlator in this sector will be of the form
\begin{displaymath}
< {\cal O}_1 \cdots {\cal O}_r > \: \sim \:
\int_{ {\cal M} } H^2\left( {\cal M}, \Lambda^2 {\cal F}^{\vee} \right)
\wedge
H^1\left( {\cal M}, {\cal F}^{\vee} \otimes {\cal F}_1 \otimes
\mbox{Obs}^{\vee} \right)
\end{displaymath}
(The fact that the first factor is degree two sheaf cohomology
is determined by a $U(1)$ selection rule, as is the fact that
the second wedge power of ${\cal F}^{\vee}$ appears, just as
in \cite{ks}.)
The factor of an element of
\begin{displaymath}
H^1\left( {\cal M}, {\cal F}^{\vee} \otimes {\cal F}_1 \otimes
\mbox{Obs}^{\vee} \right)
\end{displaymath}
reflects the contribution of the four-fermi term $F \psi \psi \lambda \lambda$
that we must necessarily use to absorb the `excess' fermi zero modes.

Note however that 
\begin{eqnarray*}
H^1\left( {\cal M}, {\cal F}^{\vee} \otimes {\cal F}_1 \otimes
\mbox{Obs}^{\vee} \right)
& = & H^1\left( {\bf P}^3, 
\oplus_1^3 {\cal O}(-2) \otimes {\cal O} \otimes {\cal O}(2) \right) \\
& = & H^1\left( {\bf P}^3, \oplus_1^3 {\cal O} \right) \\
& = & \oplus_1^3 H^1\left( {\bf P}^3, {\cal O} \right) \\
& = & 0
\end{eqnarray*}
and so any correlation function in this sector must vanish,
since the four-fermi term we must use to absorb excess fermi
zero modes, is cohomologically trivial.

Thus, for $d=1$, we see that there are no rational curve corrections
to correlation functions, precisely as expected for the B model
(but somewhat less obvious in this description as a $(0,2)$ A model
analogue).

Next we shall apply the same method to consider more general $d>0$,
and here also we shall find that there are no rational curve corrections
to correlation functions because the four-fermi terms generate a 
cohomologically-trivial factor.

A potentially nonzero correlation function is of the form
\begin{displaymath}
< {\cal O}_1 \cdots {\cal O}_r > \: \sim \:
\int_{ {\cal M} } H^{2}\left( {\cal M}, \Lambda^{2} {\cal F}^{\vee}
\right) \wedge \wedge^{2d-1} H^1\left( {\cal M}, {\cal F}^{\vee} 
\otimes {\cal F}_1 \otimes \mbox{Obs}^{\vee} \right)
\end{displaymath}
We shall argue that the group
\begin{displaymath}
H^1\left( {\cal M}, {\cal F}^{\vee} \otimes {\cal F}_1 \otimes
\mbox{Obs}^{\vee} \right)
\end{displaymath}
necessarily vanishes, hence the class in that group produced by the
four-fermi term is necessarily zero, and so
quantum corrections to correlation functions vanish.

From the long exact sequence associated to the short exact sequence
of (\ref{ff1}), we have
\begin{eqnarray}   
\lefteqn{ H^1\left( {\cal M}, {\cal F}^{\vee} \otimes {\cal O} \otimes \mbox{Obs}^{\vee}
\right) \: \longrightarrow \:
H^1\left( {\cal M}, {\cal F}^{\vee} \otimes {\cal F}_1 \otimes
\mbox{Obs}^{\vee} \right) } \nonumber\\
& &  \longrightarrow \:
H^1\left( {\cal M}, {\cal F}^{\vee} \otimes \left[ \oplus_1^2 
H^1\left( {\bf P}^1, {\cal O}(-d) \right) \otimes_{ {\bf C} }
{\cal O}(-1) \right] \otimes \mbox{Obs}^{\vee} \right)   \label{detmid}
\end{eqnarray}
It is straightforward to compute
\begin{eqnarray*}
H^1\left( {\cal M}, {\cal F}^{\vee} \otimes {\cal O} \otimes
\mbox{Obs}^{\vee} \right) & = &
H^1\left( {\bf P}^{2d+1}, \oplus_1^{(2d+1)(2d-1)} {\cal O} \right) \\
& = & \oplus_1^{(2d+1)(2d-1)} H^1\left( {\bf P}^{2d+1}, {\cal O} \right) \\
& = & 0
\end{eqnarray*}
and that
\begin{eqnarray*}
\lefteqn{ H^1\left( {\cal M}, {\cal F}^{\vee} \otimes \left[ \oplus_1^2
H^1\left( {\bf P}^1, {\cal O}(-d) \right) \otimes_{ {\bf C} }
{\cal O}(-1) \right] \otimes \mbox{Obs}^{\vee} \right) } \\
\: \: \: \: & = & H^1\left( {\cal M}, \oplus_1^{2(2d+1)(2d-1)} H^1\left(
{\bf P}^1, {\cal O}(-d) \right) \otimes_{ {\bf C} } {\cal O}(-1) \right) \\
& = & \oplus_1^N H^1\left( {\bf P}^{2d+1}, {\cal O}(-1) \right) \\
& = & 0
\end{eqnarray*}
where 
\begin{displaymath}
N \: = \: 2 (2d+1) (2d-1) h^1\left( {\bf P}^1, {\cal O}(-d) \right)
\end{displaymath}
and we have used the Bott formula \cite{okoneketal}[section I.1.1].
Thus, from the sequence (\ref{detmid}) we have that
\begin{displaymath}
H^1\left( {\cal M}, {\cal F}^{\vee} \otimes {\cal F}_1 \otimes
\mbox{Obs}^{\vee} \right) \: = \: 0
\end{displaymath}
which implies that for all $d>0$, rational curve corrections to
correlation functions must vanish, exactly as expected. 

As discussed earlier, this particular example is a bit too simplistic,
but other examples follow the same form:  quantum corrections in the (2,2)
B model whose vanishing cannot be established using purely index
theory, turn out to vanish because the natural regularization of the
two-dimensional QFT puts those corrections in the form of an 
integral over a compact moduli space of an exact form, which 
then vanishes.

\subsection{Example:  Fiber of a ruled divisor}  \label{fibrul}

Another, somewhat more complicated, example can be obtained in the
case of a Calabi-Yau containing a ruled surface.
The fiber of that ruled surface has normal bundle ${\cal O} \oplus
{\cal O}(-2)$, making maps into that fiber another situation in which
index theory arguments cannot alone explain why the B model does not
get quantum corrections.

The answer we shall find here is of the same form as for
$\widetilde{ {\bf C}^2/{\bf Z}_2}$:  a purely classical analysis says
there should be an instanton contribution of the form
$\int_{\cal M} \alpha$ for $\alpha$ some nonzero top-form on ${\cal M}$,
but when we regularize the QFT by compactifying the moduli space and extending
bundles over the compactification divisor, we shall discover that the
resulting form $\alpha$ is exact, and hence the potential contribution vanishes.

Specifically, let $X$ be the total space of the canonical bundle
over the Hirzebruch surface ${\bf F}_n$.  This space is a toric variety,
and can be described by a GLSM with chiral superfields
$s$, $t$, $u$, $v$, $p$, with charges under a pair of gauged $U(1)$'s
as given in the following table:
\begin{center}
\begin{tabular}{c|ccccc}
 & $s$ & $t$ & $u$ & $v$ & $p$ \\ \hline
$\lambda$ & $1$ & $1$ & $n$ & $0$ & $-n-2$ \\
$\mu$ & $0$ & $0$ & $1$ & $1$ & $-2$ 
\end{tabular}
\end{center}
and maps into the ${\bf P}^1$ fiber of the Hirzebruch surface
${\bf F}_n$ have degree $\vec{d} = (0,1)$.

The linear sigma model moduli space ${\cal M}$ is built from the zero
modes of the chiral superfields above, as given in the table below:
\begin{center}
\begin{tabular}{c|c}
Field & Zero modes \\ \hline
$s$, $t$ & $H^0({\bf P}^1, {\cal O}(0) ) = {\bf C}$ \\
$u$ & $H^0({\bf P}^1, {\cal O}(1)) = {\bf C}^2$ \\
$v$ & $H^0({\bf P}^1, {\cal O}(1)) = {\bf C}^2$ \\
$p$ & $H^0({\bf P}^1, {\cal O}(-2)) = 0$
\end{tabular}
\end{center}
The linear sigma model moduli space can then be described as the
quotient ${\bf C}^6//{\bf C}^{\times}\times{\bf C}^{\times}$,
in which the action on ${\bf C}^6$ is described on (homogeneous)
coordinates as follows:
\begin{center}
\begin{tabular}{c|cccccc}
 & $s_0$ & $t_0$ & $u_0$ & $u_1$ & $v_0$ & $v_1$ \\ \hline
$\lambda$ & $1$ & $1$ & $n$ & $n$ & $0$ & $0$ \\
$\mu$ & $0$ & $0$ & $1$ & $1$ & $1$ & $1$ 
\end{tabular}
\end{center}
The obstruction bundle on this moduli space is given by
\begin{displaymath}
\mbox{Obs} \: = \: 
H^1\left( {\bf P}^1, {\cal O}(-2) \right) \otimes_{ {\bf C} } 
{\cal O}(-n-2, -2) \: = \: {\cal O}(-n-2,-2)
\end{displaymath}

Following the same procedure in the previous example, we shall compute
${\cal F}$ and ${\cal F}_1$, starting with the cotangent\footnote{Note
that as the cotangent bundle is described here as the kernel
of a short exact sequence, one could write down a (0,2) GLSM describing
this bundle, and in fact one can show that that (0,2) GLSM is
nonanomalous.} bundle
\begin{displaymath}
0 \: \longrightarrow \: T^*X \: \longrightarrow \:
{\cal O}(-1,0)^2 \oplus {\cal O}(-n,-1) \oplus {\cal O}(0,-1)
\oplus {\cal O}(n+2,2) \: \longrightarrow \:
{\cal O}^2 \: \longrightarrow \: 0
\end{displaymath}
which induces
\begin{displaymath}
0 \: \longrightarrow \: {\cal F} \: \longrightarrow \:
{\cal O}(-1,0)^2 \oplus {\cal O}(n+2,2)^3 \: 
\stackrel{*}{\longrightarrow} \: {\cal O}^2 \: 
\longrightarrow \: {\cal F}_1 \: \longrightarrow \: 0
\end{displaymath}
The map $*$ is induced by the zero modes of the fields.
The factor ${\cal O}(n+2,2)^3$ is acted upon by coordinates corresponding
to $p$ zero modes, but $p$ has no zero modes, hence $*$ annihilates
${\cal O}(n+2,2)^3$.
The factor ${\cal O}(-1,0)^2$ is acted upon by coordinates corresponding to
$s$, $t$ zero modes.  Both of those fields have a single zero mode, but
the structure of the map is such that all of ${\cal O}(-1,0)$ is
mapped to only one of the pair ${\cal O}^2$.

Thus, the induced long exact sequence breaks up into a pair of short
exact sequences
\begin{displaymath}
\begin{array}{c}
0 \: \longrightarrow \: {\cal F} \: \longrightarrow \:
{\cal O}(-1,0)^2 \oplus {\cal O}(n+2,2)^3 \: \longrightarrow \: {\cal O}
\: \longrightarrow \: 0 \\
{\cal F}_1 \: \cong \: {\cal O}
\end{array}
\end{displaymath}

Finally, we need to calculate the group
$H^1\left({\cal M}, {\cal F}^{\vee} \otimes {\cal F}_1 \otimes
\mbox{Obs}^{\vee} \right)$. 
Tensoring the short exact sequence for ${\cal F}^{\vee}$
\begin{displaymath}
0 \: \longrightarrow \: {\cal O} \: \longrightarrow \:
{\cal O}(1,0)^2 \oplus {\cal O}(-n-2,-2)^2 \: \longrightarrow \:
{\cal F}^{\vee} \: \longrightarrow \: 0
\end{displaymath}
with ${\cal F}_1 \otimes \mbox{Obs}^{\vee}$ we find
\begin{displaymath}
0 \: \longrightarrow \: {\cal O}(n+2,2) \: \longrightarrow \:
{\cal O}(n+3,2)^2 \oplus {\cal O}^2 \: \longrightarrow \:
{\cal F}^{\vee} \otimes {\cal F}_1 \otimes \mbox{Obs}^{\vee}
\: \longrightarrow \: 0
\end{displaymath}
which induces a long exact sequence
\begin{displaymath}
\cdots \: \longrightarrow \: H^1\left( {\cal M},
{\cal O}(3,2)^2 \oplus {\cal O}^2 \right) \: \longrightarrow \:
H^1\left({\cal M}, {\cal F}^{\vee} \otimes {\cal F}_1 \otimes
\mbox{Obs}^{\vee} \right) \: \longrightarrow \:
H^2\left({\cal M}, {\cal O}(2,2) \right) \: \longrightarrow \: \cdots
\end{displaymath}

In the case $n=0$, so that ${\cal M} = {\bf P}^1 \times {\bf P}^3$,
it is straightforward to compute that
\begin{displaymath}
H^1\left( {\cal O}(3,2)^2 \oplus {\cal O}^2 \right) \: = \: 0 \: = \:
H^2\left( {\cal M}, {\cal O}(2,2) \right)
\end{displaymath}
from which we conclude that $H^1({\cal M}, {\cal F}^{\vee} \otimes
{\cal F}_1 \otimes \mbox{Obs}^{\vee}) = 0$, which means that
any correlation functions in this sector must vanish.

Thus, once again we see that the B model is classical, despite not
appearing so immediately, because the natural regularization of the QFT
extends bundles over the compactification of the moduli space in such a way
as to make quantum corrections proportional to the integral of an exact
form.

\section{Relation between the A and B models}
\label{relnAB}

\subsection{A puzzle and its resolution}

As described earlier, classically the (0,2) A and B model analogues are
related by exchanging the gauge bundle with its dual:  
${\cal E} \mapsto {\cal E}^{\vee}$.  In this section we will argue that
there are also choices of regularizations of the two-dimensional quantum
field theory which must be exchanged at the same time.

In particular, on the basis of the classical result along,
in the special case that the pullback of the tangent
bundle to the worldsheet is isomorphic to its dual, we can formulate
an apparent contradiction.  The resolution of that contradiction will
involve understanding regularizations of the two theories.

Consider the ordinary (2,2) A model, in a case in which it receives
quantum corrections from curves such that $\phi^* T \cong \phi^* T^{\vee}$.
Then, using the advertised relation between the A and B models,
\begin{eqnarray*}
\mbox{contribution to (2,2) A model} & = &
\mbox{contribution to (0,2) B model with } {\cal E} = T^{\vee} \\
& = & \mbox{contribution to (0,2) B model with } {\cal E} = T \\
& & \mbox{ since } {\cal E} \cong {\cal E}^{\vee} \\
& = & \mbox{contribution to the (2,2) B model}
\end{eqnarray*}
But the B model receives no quantum corrections, so if such an example
exist, we appear to have a problem with our claimed relation between
the A and B models.

Such examples certainly exist.  For example, if a Calabi-Yau threefold contains
a divisor corresponding to a Hirzebruch surface, then rational
curves corresponding to the ${\bf P}^1$ fiber of the Hirzebruch surface
form one example.
Perhaps the easiest such example is
the space $\widetilde{ {\bf C}^2/{\bf Z}_2 }$.
The A model (uncoupled to topological gravity) does receive quantum corrections;
however, since $\phi^* T \cong {\cal O}(2) \oplus {\cal O}(-2)$, we see
that $\phi^* T \cong \phi^* T^{\vee}$, putting us precisely in the situation
above which apparently yields a contradiction.

To understand the resolution of this puzzle, let us study the details of
the calculations.
For simplicity, let us consider the case of $\widetilde{ {\bf C}^2/{\bf Z}_2 }$.
(Strictly speaking, K3's do not receive quantum corrections, but 
because cohomology classes on the K3 induce vanishing cohomology 
on ${\cal M}$ -- otherwise, the mechanical details of the calculations
form the simplest possible example of the phenomenon under discussion.
Thus, we shall describe calculations on this space, as the results
form the prototype for more complicated cases.)
In the case of the ordinary A model on this space, the linear
sigma model moduli space ${\cal M} = {\bf P}^{2d + 1}$ for degree $d$
maps, as demonstrated earlier in section~\ref{c2z2B}.
On the (2,2) locus, with the physically-canonical representation of the
tangent bundle, it is straightforward to compute
\cite{ks}
that for ${\cal E} = TX$, $X = \widetilde{ {\bf C}^2/{\bf Z}_2}$,
${\cal F} = T {\cal M}$ and ${\cal F}_1$ is the obstruction bundle,
\begin{displaymath}
\mbox{Obs} \: = \: H^1( {\bf P}^1, {\cal O}(-2d) ) \otimes_{ {\bf C} }
{\cal O}(-2)
\end{displaymath}
which has rank $2d-1$.

By contrast, when we take ${\cal E}= (TX)^{\vee}$ and compute the
in the (0,2) analogue of the B model, we found in section~\ref{c2z2B} that
\begin{displaymath}
\begin{array}{c}
{\cal F}_B \: = \: H^0( {\bf P}^1, {\cal O}(2d) ) \otimes_{ {\bf C}} {\cal O}(2)
\\
0 \: \longrightarrow \: {\cal F}_{B1} \: \longrightarrow \:
\left[ H^1({\bf P}^1, {\cal O}(-d) ) \otimes_{ {\bf C} } {\cal O}(-1) 
\right]^{\oplus 2} \: \longrightarrow \: 0
\end{array}
\end{displaymath}

Now, recall that for honest maps,
\begin{eqnarray*}
{\cal F} & = & R^0 \pi_* \alpha^* {\cal E} \\
{\cal F}_B & = & R^0 \pi_* \alpha^* ( {\cal E}^{\vee} ) \\
{\cal F}_1 & = & R^1 \pi_* \alpha^* {\cal E} \\
{\cal F}_{B1} & = & R^1 \pi_* \alpha^* ( {\cal E}^{\vee} )
\end{eqnarray*}
so given the fact that for all honest maps, $\phi^* {\cal E} \cong
\phi^* {\cal E}^{\vee}$ here, one would expect that
${\cal F} \cong F_B$ and ${\cal F}_1 \cong {\cal F}_{B1}$.

However, that is not what we see above.
The rank of ${\cal F}$ matches that of ${\cal F}_B$, and the rank of
${\cal F}_1$ matches that of ${\cal F}_{B1}$, but otherwise they are
distinct bundles.  

In particular, they are distinct because of their extensions over the
compactification divisor -- on the interior of the moduli space,
corresponding to honest maps, they are isomorphic.
In our previous work \cite{ks}, we gave a general argument for why
the bundles created using the LSM-based ansatz match $R^i \pi_*
\alpha^* {\cal E}$ on the interior of the moduli space; in the present
case, since ${\cal F}$, ${\cal F}_B$, for example, both correspond to
extensions of the same $R^0 \pi_* \alpha^*$, they must be the same
on the interior.  For $d=1$ maps it is very easy to see this result.
The compactified moduli space is ${\cal M} = {\bf P}^3$, and the
space of honest maps is given by $SL(2,{\bf C}) \subset {\bf P}^3$.
We found above that ${\cal F} = T {\cal M}$ and ${\cal F}_B = {\cal O}(2)^3$.
However, over $SL(2,{\bf C})$, the only\footnote{The group manifold
$SL(2,{\bf C})$ is homotopic to $S^3$, for which $H^2 = 0$.} 
line bundle is the trivial line
bundle, and since it is a group manifold, its tangent bundle is trivializable.
Thus, when we restrict to the interior of the moduli space, we find that
\begin{displaymath}
T {\cal M}_{int} \: \cong \: {\cal O}^3 \: \cong \:
{\cal O}(2)|_{int} 
\end{displaymath}
and so in particular ${\cal F}|_{int} \cong {\cal F}_B|_{int}$.
Similarly, ${\cal F}_1|_{int} \cong {\cal F}_{B 1}|_{int}$.
Although the bundles are different over the compactification divisor,
on the interior of the moduli space they are the same.

The analysis of the example in section~\ref{fibrul} is similar but
with a slightly different conclusion in that case.
There, for $n=0$, ${\cal M} = {\bf P}^1 \times {\bf P}^3$.
For the A model, ${\cal F} \cong T {\cal M}$, and ${\cal F}_1  \cong
\mbox{Obs} \cong {\cal O}(-n-2,2)$, whereas for the B model,
\begin{displaymath}
\begin{array}{c}
0 \: \longrightarrow \: {\cal F}_B \: \longrightarrow \:
{\cal O}(-1,0)^2 \oplus {\cal O}(2,2)^3 \: \longrightarrow \: 
{\cal O} \: \longrightarrow \: 0 \\
{\cal F}_{B 1} \: \cong \: {\cal O}
\end{array}
\end{displaymath}
Here, the interior of the moduli space is given by ${\bf P}^1 
\times SL(2,{\bf C})$, and when we restrict to that interior, we find
that the restrictions of ${\cal F}$ and ${\cal F}_B$ to the interior are
{\it not} the same:  one looks like $T {\bf P}^1$ over the
${\bf P}^1$ whereas the other looks like $T^* {\bf P}^1$ over the
${\bf P}^1$.

In this case as before, $\phi^* {\cal E} \cong \phi^* {\cal E}^{\vee}$
for any $\phi$, but here the choice of isomorphism fibers together nontrivially
over the moduli space, so that the induced sheaves are {\it not} isomorphic
on the interior of the moduli space, unlike the last case.

As a result, the naive contradiction can be resolved in a much easier way
in this case than the last:  here, the classical contributions to the
A and B models are different, whereas previously they were the same.

In short, in these two examples we have found two different mechanisms
by which physics resolves the naive contradiction generated in the
beginning.  In the second case, although the contributions from
any single instanton naively appear as if they should be the same,
those contributions fit together differently, giving different induced
sheaves even over the interior of the moduli space describing honest
maps.  In the first case, on the other hand, the induced sheaves
are isomorphic on the interior of the moduli space, but they differ
over the compactification divisor -- there are two distinct regularizations
of the physical theory.

As a further check, let us compare four-fermi terms.
In the A model in the first case, the four-fermi terms are interpreted
as generating factors of elements of
\begin{displaymath}
H^1\left( {\cal M}, {\cal F}^{\vee} \otimes {\cal F}_1 \otimes
\mbox{Obs}^{\vee} \right) \: = \: H^1\left( {\cal M},
(T {\cal M})^{\vee} \otimes \mbox{End } \mbox{Obs} \right)
\end{displaymath}
Here, $\mbox{End }\mbox{Obs} = \mbox{Obs}^{\vee} \otimes \mbox{Obs}$
is isomorphic to $h^1({\bf P}^1, {\cal O}(-2d))^2 = (2d-1)^2$ copies
of the trivial line bundle ${\cal O}$.
It is a standard result that $h^1( {\bf P}^{2d+1}, \Omega^1 ) = 1$.
By contrast, in our (0,2) B model computation in section~\ref{c2z2B},
we argued that the sheaf cohomology group that the four-fermi term lives
in, namely
\begin{displaymath}
H^1\left( {\cal M}, {\cal F}_B^{\vee} \otimes {\cal F}_{B1} \otimes
\mbox{Obs}^{\vee} \right)
\end{displaymath}
necessarily vanishes.
This difference between the four-fermi terms is the mechanical reason
why the A model gets corrections in this case and the B model does not:
any quantum corrections necessarily involve bringing down factors of the
four-fermi term; such factors are nonvanishing only for the A model.

To summarize, to say that to exchange the (0,2) analogues of the A and B models
we merely need exchange ${\cal E}$ with ${\cal E}^{\vee}$ is a bit
naive -- more precisely, in general we must also exchange two different
regularizations at the same time.
As usual, the gauged linear sigma model seems to know about this fact
implicitly, as using GLSM-based regularizations automatically produces
correct physical results.

\subsection{Speculations on auxiliary fields}

We have learned that to understand at a mechanical level why
the B model is classical, often requires one to understand how the theory
is regularized, and suitable compactifications of moduli spaces and
extensions of sheaves that define that regularization.

This result is somewhat reminiscent of old results concerning contact
terms and auxiliary fields \cite{greenseiberg}.  
It is an old result that to understand
contact terms involves understanding the behavior of a theory
on the compactification divisor of a moduli space of marked Riemann
surfaces, and furthermore that the necessity of contact terms
in at least some situations stems from working on-shell, where
auxiliary fields have been integrated out -- working with multiplets
containing auxiliary fields sometimes removes the need for
contact terms.

In the present case, one might wonder if there is an analogous
phenomenon.  Our moduli spaces are not moduli spaces of marked
Riemann surfaces, but rather moduli spaces of parametrized maps,
so compactifying the moduli space does not give information
about contact term interactions.
On the other hand, we have described the twists of the (0,2) nonlinear
sigma model on-shell; if one works off-shell, then one finds
that the auxiliary fields are not symmetric under the
${\cal E} \leftrightarrow {\cal E}^{\vee}$ interchange,
not even classically. 

To push the analogy to \cite{greenseiberg}, one would want
a mechanism that directly relates compactifications of moduli spaces
(not necessarily of marked Riemann surfaces) to auxiliary fields.
Such an understanding of \cite{greenseiberg} was given in
\cite{distlerdoyle}.  Briefly, there it was argued that the
contact terms of \cite{greenseiberg} could alternatively be
understood in terms of a non-split supermoduli space.
The authors of \cite{distlerdoyle} argued that a more complete
understanding of the quantum field theory in question
leads one to understand the zero modes of the theory in terms
of a supermoduli space, instead of a moduli space, and 
the presence of auxiliary fields implies that the supermoduli space
does not split holomorphically.  Taking into account the effects
of that nonsplit supermoduli space leads to the same results
as in \cite{greenseiberg}.

In the present case of the (0,2) analogues of the A, B models,
one might therefore speculate that the zero modes of the two-dimensional
quantum field theory should be understood in terms of a 
(possibly nonsplit) supermoduli space.  Perhaps the effects of
four-fermi terms can be understood in terms of integrating over a 
nonsplit supermoduli space, for example.

Another possibility is suggested by \cite{distnel}, which 
related topological gravity and its cohomological-field-theory aspects
to local superconformal symmetry. 

We leave these lines of thought for future work.

\section{Hints of a nonrenormalization result}
\label{nonrenorm}

In this section we shall consider another example
of a (0,2) B model computation, viewed as a (0,2) A model computation,
in which we take the gauge bundle to be a deformation of the 
cotangent bundle.  We shall find that the (0,2) B model correlation
functions vanish not only on the (2,2) locus, but also off the (2,2)
locus as well, suggesting there is an applicable nonrenormalization result.

Our toric Calabi-Yau is the total space of the canonical bundle to
${\bf P}^1 \times {\bf P}^1$, which can be described by a gauged
linear sigma model with fields $x_1$, $x_2$, $\tilde{x}_1$,
$\tilde{x}_2$, $p$, with charges under a pair of gauged $U(1)$'s as
\begin{center}
\begin{tabular}{c|ccccc}
 & $x_1$ & $x_2$ & $\tilde{x}_1$ & $\tilde{x}_2$ & $p$ \\ \hline
$\lambda$ & $1$ & $1$ & $0$ & $0$ & $-2$ \\
$\mu$ & $0$ & $0$ & $1$ & $1$ & $-2$ \\
\end{tabular}
\end{center}
The tangent bundle $T$ to this Calabi-Yau is the cokernel of a short
exact sequence
\begin{displaymath}
0 \: \longrightarrow \: {\cal O}^2 \: \stackrel{A}{\longrightarrow} \:
{\cal O}(1,0)^2 \oplus {\cal O}(0,1)^2 \oplus {\cal O}(-2,-2) \:
\longrightarrow \: T \: \longrightarrow \: 0
\end{displaymath}
where the map $A$ is given by
\begin{displaymath}
\left[ \begin{array}{cc}
       x_1 & 0 \\
       x_2 & 0 \\
        0 & \tilde{x}_1 \\
        0 & \tilde{x}_2 \\
       -2p & -2p
       \end{array} \right]
\end{displaymath}

We shall compute quantum corrections to correlation functions in 
a (0,2) deformation of the B model on the Calabi-Yau above,
as a (0,2) A model computation with gauge bundle a deformation of
the cotangent bundle.  This gauge bundle ${\cal E}$ can be described as
the kernel of the short exact sequence
\begin{displaymath}
0 \: \longrightarrow \: {\cal E} \: \longrightarrow \:
{\cal O}(-1,0)^2 \oplus {\cal O}(0,-1)^2 \oplus {\cal O}(2,2) \:
\stackrel{B}{\longrightarrow} \: {\cal O}^2 \: \longrightarrow \: 0
\end{displaymath}
where the map $B$ is a deformation of the dual of the map $A$,
specifically
\begin{displaymath}
\left[ \begin{array}{ccccc}
       x_1 & x_2 & \beta_1 \tilde{x}_1 + \beta_2 \tilde{x}_2 &
       \beta_1' \tilde{x}_1 + \beta_2' \tilde{x}_2 & -2p \\
       \alpha_1 x_1 + \alpha_2 x_2 & \alpha_1' x_1 + \alpha_2' x_2 &
       \tilde{x}_1 & \tilde{x}_2 & -2p
       \end{array}
       \right]
\end{displaymath}
where $\alpha_i$, $\alpha_i'$, $\beta_i$, $\beta_i'$ are constants.

Let $d_1$, $d_2$ denote the worldsheet instanton numbers with respect
to the gauged $U(1)$'s $\lambda$, $\mu$, respectively.
The normal bundle\footnote{To give some idea of where this comes from,
the $2 d_1 d_2$ is the normal bundle in ${\bf P}^1 \times {\bf P}^1$
computed as the self-intersection $(d_1 H_1 + d_2 H_2)^2 = 2 d_1 d_2$,
where the $H_i$ are the hyperplane (point) classes in the
respective ${\bf P}^1$'s.} to such a rational curve is given by
\begin{displaymath}
{\cal O}(2 d_1 d_2) \oplus {\cal O}(-2 d_1 d_2 - 2)
\end{displaymath}
so from our general analysis earlier we can already see that
correlation functions can only possibly receive contributions
from sectors in which one of $d_1$, $d_2$ is one and the other zero,
but for completeness we shall work through all the details regardless.

For the moment, let us assume both $d_1 > 0$ and $d_2 > 0$.
As mentioned above, from general analysis there should not be any
contributions, but let us work through the details regardless.
The linear sigma model moduli space is easily computed to be
\begin{displaymath}
{\cal M} \: = \: {\bf P}^{2(d_1+1)-1} \times
{\bf P}^{2(d_2+1)-1}
\end{displaymath}
and the obstruction sheaf is given by
\begin{displaymath}
\mbox{Obs} \: = \: H^1\left( {\bf P}^1, {\cal O}(-2 d_1 -2 d_2) \right)
\otimes_{ {\bf C} } {\cal O}(-2,-2)
\end{displaymath}
The induced sheaves ${\cal F}$, ${\cal F}_1$ are defined by the
long exact sequence
\begin{displaymath}
\begin{array}{ccccc}
0 & \longrightarrow & {\cal F} & \longrightarrow &
\oplus_1^2 \left[ H^0\left( {\bf P}^1, {\cal O}(-d_1) \right) \otimes_{ {\bf C}}{\cal O}(-1,0) \right]
\oplus_1^2 \left[ H^0\left( {\bf P}^1, {\cal O}(-d_2) \right) \otimes_{ {\bf C}}{\cal O}(0,-1) \right] \\
 & & & & \oplus H^0\left( {\bf P}^1, {\cal O}(2d_1 + 2d_2) \right)\otimes_{ {\bf C}}
{\cal O}(2,2) \: \stackrel{*}{\longrightarrow} \: {\cal O}^2\\
& \longrightarrow & {\cal F}_1 & \longrightarrow &
\oplus_1^2 \left[ H^1\left( {\bf P}^1, {\cal O}(-d_1) \right) \otimes_{ {\bf C}}{\cal O}(-1,0) \right]
\oplus_1^2 \left[ H^1\left( {\bf P}^1, {\cal O}(-d_2) \right) \otimes_{ {\bf C}}{\cal O}(0,-1) \right] \\
& & & & \oplus H^1\left( {\bf P}^1, {\cal O}(2d_1 + 2d_2) \right)\otimes_{ {\bf C}}
{\cal O}(2,2) \: \longrightarrow \: 0
\end{array}
\end{displaymath}
Let us consider for a moment the map $*$.
This map is induced by expanding the map $B$ above in the zero modes
of its constituents.  However, most of what $*$ acts on is zero.
The only nonzero part that $*$ acts on, is multiplied by zero modes
of $p$, but $p$ has no zero modes.  Thus, the map $*$ is identically zero.
As a result, we can simplify the long exact sequence above to 
\begin{displaymath}
\begin{array}{c}
{\cal F} \: \cong \: H^0\left( {\bf P}^1, {\cal O}(2 d_1 + 2 d_2) \right)
\otimes_{ {\bf C}} {\cal O}(2,2) \\
0 \: \longrightarrow \: {\cal O}^2 \: \longrightarrow \:
{\cal F}_1 \: \longrightarrow \: 
\oplus_1^{2(d_1-1) }{\cal O}(-1,0) \oplus_1^{2(d_2-1)} {\cal O}(0,-1)
\: \longrightarrow \: 0
\end{array}
\end{displaymath}
from which we see that
\begin{eqnarray*}
\mbox{rank }{\cal F} & = & 2(d_1 + d_2) + 1 \\
\mbox{rank }{\cal F}_1 & = & 2(d_1 + d_2) - 2
\end{eqnarray*}
However, note that the rank of ${\cal F}_1$ does not match that
of $\mbox{Obs}$, and so for the reasons described earier,
there can be no contribution to correlation functions from this
instanton sector.

The cases that might contribute are when $d_1=1$ and $d_2=0$,
and when $d_1=0$ and $d_2=1$.
As these are symmetric, we shall examine only the former.

Here, the linear sigma model moduli space is ${\bf P}^3 \times {\bf P}^1$,
and the obstruction bundle is ${\cal O}(-2,-2)$.
The analysis proceeds almost exactly as above, 
except that the map $*$ will no longer be identically zero:
the components
\begin{displaymath}
\oplus_1^2 H^0\left( {\bf P}^1, {\cal O}(-d_2) \right)\otimes_{ {\bf C}}
{\cal O}(0,-1)
\end{displaymath}
are nonzero, and are multiplied by nonzero parts of the zero mode expansion
of the map $B$.
As a result, the map $*$ will be onto ${\cal O}$ if all the $\beta_i$
and $\beta_i'$ vanish, and onto ${\cal O}^2$ more generally.

Suppose the map $*$ maps onto ${\cal O}^2$.
Then, we have
\begin{displaymath}
\begin{array}{c}
0 \: \longrightarrow \: {\cal F} \: \longrightarrow \:
{\cal O}(0,-1)^2 \oplus_1^3{\cal O}(2,2) \: \longrightarrow \:
{\cal O}^2 \: \longrightarrow \: 0 \\
{\cal F}_1 \: \cong \: 0
\end{array}
\end{displaymath}
In this case, ${\cal F}_1$ has rank $0$, but the obstruction sheaf
has rank one, so again there is a mismatch of excess fermi zero modes,
four-fermi terms cannot absorb the excess zero modes,
and so there can be no contribution to a correlation function.

Suppose instead that $*$ maps onto ${\cal O}$.
Then we have
\begin{displaymath}
\begin{array}{c}
0 \: \longrightarrow \: {\cal F} \: \longrightarrow \:
{\cal O}(0,-1)^2 \oplus_1^3{\cal O}(2,2) \: \longrightarrow \:
{\cal O} \: \longrightarrow \: 0 \\
{\cal F}_1 \: \cong \: {\cal O}
\end{array}
\end{displaymath}
Now, the rank of ${\cal F}_1$ does match the rank of the obstruction
bundle, so the excess fermi zero modes do match, and it is
potentially possible to get a nonzero correlation function via
four-fermi terms.  Note that this would be the only case in which
there could be an instanton contribution to a B-analogue coupling
in this theory.

However, let us examine this case a little more closely.
We shall see that the group
\begin{displaymath}
H^1\left( {\cal M}, {\cal F}^{\vee} \otimes {\cal F}_1 \otimes
\mbox{Obs}^{\vee} \right)
\end{displaymath}
vanishes, so there can be no instanton contribution in this case.
Indeed, this had better be the case:  the case that the
map $*$ is onto, namely when the $\beta_i$ and $\beta_i'$ vanish,
includes the (2,2) locus as a special case.  We know that there are no
instanton corrections to the (2,2) locus, so as an instanton contribution
here would imply a contribution to the (2,2) locus, we know that there
had better not be any instanton contribution from this sector.

Before calculating the group above, let us collect some
pertinent facts.  The linear sigma model moduli space in this
case is ${\bf P}^3 \times {\bf P}^1$, the obstruction sheaf is
${\cal O}(-2,-2)$, ${\cal F}_1 \cong {\cal O}$, and
\begin{displaymath}
0 \: \longrightarrow \: {\cal F} \: \longrightarrow \:
{\cal O}(0,-1)^2 \oplus_1^3{\cal O}(2,2) \: \longrightarrow \:
{\cal O} \: \longrightarrow \: 0
\end{displaymath}
From that short exact sequence we have a long exact sequence with terms
\begin{displaymath}
H^1\left( {\cal M}, \left[ {\cal O}(0,1)^2 \oplus_1^3{\cal O}(-2,-2)
\right] \otimes {\cal O}(2,2) \right) \: \longrightarrow \:
H^1\left( {\cal M}, {\cal F}^{\vee} \otimes {\cal F}_1 \otimes
\mbox{Obs}^{\vee} \right) 
\: \longrightarrow \:
H^2\left( {\cal M}, 
{\cal O}(2,2) \right)
\end{displaymath}
Now we need merely evaluate the terms in this sequence.
First, we know
\begin{displaymath}
H^2\left( {\cal M}, {\cal O}(2,2) \right) \: = \:
H^0\left( {\bf P}^1, {\cal O}(2) \right) \otimes
H^2\left( {\bf P}^3, {\cal O}(2) \right)
\end{displaymath}
but from Bott's formula \cite{okoneketal}[section I.1.1]
the second factor vanishes.
Similarly,
\begin{displaymath}
H^1\left( {\cal M}, {\cal O}(2,3)^2 \oplus_1^3 {\cal O} \right) \: = \:
\oplus_1^2 \left[ H^1\left( {\cal M}, {\cal O}(2,3) \right) \right]
\oplus_1^3 H^1\left( {\cal M}, {\cal O} \right)
\end{displaymath}
but using Bott's formula and proceeding as above we find that
\begin{displaymath}
H^1\left( {\cal M}, {\cal O}(2,3) \right) \: = \: 0 \: = \:
H^1\left( {\cal M}, {\cal O} \right)
\end{displaymath}
so as all other terms in the long exact sequence vanish,
we find that
\begin{displaymath}
H^1\left({\cal M}, {\cal F}^{\vee} \otimes {\cal F}_1 \otimes
\mbox{Obs}^{\vee} \right) \: = \: 0
\end{displaymath}

Thus, in this example of a family of B model analogues,
quantum corrections to correlation functions vanish
not only on the (2,2) locus but also off the (2,2) locus,
suggesting the existence of a nonrenormalization theorem.

In fact, one can outline situations in which such
nonrenormalization results would hold.
Suppose on the (2,2) locus the rational curves all have a normal
bundle with the property that B model contributions are excluded
by purely index theory.  Then, if the deformation off the (2,2) locus
is constrained such that the tangent bundle of the worldsheet is
a subbundle for all curves, then the splitting type of the pullback
to the worldsheet will not change, and so the same index theory argument
that precluded quantum corrections on the (2,2) locus, will also
apply off the (2,2) locus.

\section{Conclusions}

In this paper we have extended previous work \cite{ks} on the
(0,2) analogue of the A model to the (0,2) analogue of the B model.
Classically, the (0,2) A and B models are related by 
the exchange of the gauge bundle with its dual, but quantum-mechanically
one discovers that one must often regularize the theories in two different
ways, which resolves some naive contradictions in the resulting
picture.

Along the way, we have also learned that the ordinary (2,2) B model
can be described on slightly more spaces than just Calabi-Yau's:
one merely needs to require that $K^{\otimes}$ be trivial.

One direction that needs to be better understood is the behavior
of (0,2) quantum cohomology rings through the bundle analogue of
flops.  Some bundles are only stable in a subset of the K\"ahler cone;
the K\"ahler cone breaks up into subcones, with a different moduli space
of bundles in each subcone.  This phenomenon was described in detail
in \cite{kcsub}, and describes a bundle analogue of flops.
Ordinary (2,2) quantum cohomology rings of Calabi-Yau manifolds
are invariant under flops, so it is natural to ask whether
(0,2) quantum cohomology rings on Calabi-Yau's are invariant under
these bundle analogues of flops.

\section{Acknowledgements}

We would like to thank A.~Adams, R.~Donagi, M.~Gross,
S.~Hellerman, D.~Morrison, and especially J.~Distler and S.~Katz
for useful conversations.

\appendix

\section{Consistency conditions in the closed string $(2,2)$ B model}   
\label{bmod}

In this appendix we will argue that for the
closed string B model, the Calabi-Yau condition can be weakened
very slightly, to the constraint that the canonical divisor
be 2-torsion.
This proposed weakening is specific to the closed string B model;
consistency conditions for the open string B model still require
that the target space be Calabi-Yau.

\subsection{Anomalies}

Before discussing the anomaly cancellation condition in the B model,
let us take a moment to review the A model, which does not suffer
from any anomalies on non-Calabi-Yau complex K\"ahler manifolds.
The worldsheet fermions in the closed string A model are defined
as follows \cite{edtft,ks}:
\begin{eqnarray*}
\psi_+^i & \in & \Gamma_{ C^{\infty} }\left( \phi^* T^{1,0}X \right) \\
\psi_+^{\overline{\imath}} & \in &
\Gamma_{ C^{\infty} }\left( K_{\Sigma} \otimes \left(
\phi^* T^{1,0}X \right)^{\vee} \right) \\
\psi_-^i & \in & \Gamma_{ C^{\infty} }\left(
\overline{K}_{\Sigma} \otimes \left( \phi^* T^{0.1}X \right)^{\vee} \right) \\
\psi_-^{\overline{\imath}} & \in &
\Gamma_{ C^{\infty} }\left( \phi^* T^{0,1}X \right)
\end{eqnarray*}
Here, classically, when the worldsheet is ${\bf P}^1$ it is the
$\psi_+^i$ and $\psi_-^{\overline{\imath}}$ that have zero modes,
as many zero modes as the complex dimension of $X$.
Naively, in order for the path integral measure to be well-defined,
one might think that one needs separately a holomorphic nowhere-zero section of
$\Lambda^{top} T^{1,0}X$ and an antiholomorphic nowhere-zero section
of
$\Lambda^{top} T^{0,1}X$, which would contradict known
results about the A model.  

However,
this is stronger than we need.  All we need is for the tensor product
of the left- and right-moving zero-mode bundles to be topologically
trivial, together with a nowhere-zero holomorphic section of the
tensor product.  (Holomorphic because we want the section to be
invariant under the right-moving supersymmetry -- 
lack of holomorphicity would spontaneously break supersymmetry.)

In other words,
the resolution of our apparent paradox is that we do not need to have
nowhere-zero sections of the two bundles separately, we really only
need to have a section of the product of the two
in order for the
path integral measure to be well-defined.
All we really need is a holomorphic nowhere-zero section of
the product
\begin{displaymath}
\left( \Lambda^{top} T^{1,0} X \right) \otimes
\left( \Lambda^{top} T^{1,0}X \right)^{\vee} 
\end{displaymath}
The metric on $X$ induces an isomorphism between
$\Lambda^{top}
T^{0,1}X$ and $\Lambda^{top} (T^{1,0}X)^{\vee}$ as topological bundles,
so
\begin{displaymath}
\left( \Lambda^{top} T^{1,0} X \right) \otimes
\left( \Lambda^{top} T^{1,0}X \right)^{\vee} \: \cong \: {\cal O}_X
\end{displaymath}
A holomorphic nowhere-zero section of this bundle
always exists, regardless of $X$,
even when sections of the individual factors do not
exist separately.
Thus, we
recover the well-known fact that the A model can be defined on more
spaces than just Calabi-Yau's.
(We have only discussed the classical sector of the A model,
but in sections~\ref{anomg0}, \ref{anomhg} we shall repeat this anomaly analysis in quantum sectors
of a generalization of the A model, and there we will see that the
same result holds.)

Now that we have reviewed anomaly cancellation in the A model,
let us apply the same analysis to the closed string B model.
The worldsheet fermions in the closed string B model
are defined as follows \cite{edtft,ks}:
\begin{eqnarray*}
\psi_+^i & \in & \Gamma_{ C^{\infty} }\left( \phi^* T^{1,0}X \right) \\
\psi_+^{\overline{\imath}} & \in &
\Gamma_{ C^{\infty} }\left( K_{\Sigma} \otimes \left(
\phi^* T^{1,0}X \right)^{\vee} \right) \\
\psi_-^i & \in & \Gamma_{ C^{\infty} }\left(
\left( \phi^* T^{0.1}X \right)^{\vee} \right) \\
\psi_-^{\overline{\imath}} & \in &
\Gamma_{ C^{\infty} }\left( \overline{K}_{\Sigma} \otimes
\phi^* T^{0,1}X \right)
\end{eqnarray*}
The zero modes of the $\psi_+$ are holomorphic sections of the
given bundles, and the zero modes of the $\psi_-$ are antiholomorphic
sections of the given bundles.
When the worldsheet is ${\bf P}^1$, only $\psi_+^i$ and $\psi_-^i$
have zero modes, and each has as many zero modes as the complex
dimension of $X$.  Thus, in order for the path integral measure to
be well-defined, one would naively think that
we need the product of a holomorphic nowhere-zero section of
$\Lambda^{top} T^{1,0}X$ and an antiholomorphic nowhere-zero section
of $\Lambda^{top} (T^{0,1}X)^{\vee}$, which would lead one to the usual
conclusion that $X$ must be Calabi-Yau in order for the B model to
be well-defined.  However, in our analysis of the A model we saw
that such a conclusion was too strong; in order to make the path integral
measure well-defined, one needs only a section of the product of
the bundles.
Just as before, we can use the Riemannian metric to map
the antiholomorphic vector bundle $\Lambda^{top} (T^{0,1}X)^{\vee}$
to the holomorphic vector bundle $\Lambda^{top} T^{1,0}X$, as topological
bundles.
As a result, just as
in our analysis of the A model,
instead of demanding separately a holomorphic nowhere-zero section
of $\Lambda^{top} T^{1,0}X$ and an antiholomorphic nowhere-zero section
of $\Lambda^{top} (T^{0,1}X)^{\vee}$, it suffices to have a holomorphic
nowhere-zero section of
\begin{displaymath}
\left( \Lambda^{top} T^{1,0} X \right) \otimes
\left( \Lambda^{top} T^{1,0} X \right)
\end{displaymath}
which is equivalent to the constraint that $K_X^{\otimes 2} \cong
{\cal O}_X$.

Thus, repeating our analysis of the A model in the case of the B model,
we find that on genus zero worldsheets
the Calabi-Yau condition can be very slightly weakened;
we only seem to need $K_X^{\otimes 2} \cong {\cal O}_X$,
and not necessarily $K_X \cong {\cal O}_X$.

So far we have discussed genus zero worldsheets, but it is easy
to extend the result to higher genus worldsheets of fixed\footnote{We
are {\it not} coupling to topological gravity in this paper, for reasons
partly discussed previously in \cite{ks}.} complex
structure.
In the case of the A model, classically, the $\psi_+^i$ zero modes
couple to $h^0(\Sigma, {\cal O})$ copies of $\phi^* T^{1,0}X$,
whereas the $\psi_+^{\overline{\imath}}$ zero modes couple to
$h^1(\Sigma, {\cal O})$ copies of $\phi^* T^{1,0}X$, and similarly
for the left-movers.  As a result, for the $(2,2)$ A model
path integral measure to be
well-defined, one needs a holomorphic nowhere-zero section of
\begin{displaymath}
\left( \Lambda^{top} T^{1,0} X \right)^{\chi/2} \otimes
\left( \Lambda^{top} T^{1,0} X \right)^{-\chi/2} \: \cong \:
{\cal O}_X
\end{displaymath}
and as such a section always exists, there is no constraint.
We have implicitly used the fact that
\begin{displaymath}
h^0\left(\Sigma, {\cal O}\right) \: - \:
h^1\left( \Sigma, {\cal O}\right) \: = \: \chi(\Sigma, {\cal O}) \: = \:
1 - g \: = \: \chi/2
\end{displaymath}
The analysis of the $(2,2)$ closed string B model at genus $g$ is very
similar, proceeding as above one finds that one needs a holomorphic
nowhere-zero section of
\begin{displaymath}
\left( \Lambda^{top} T^{1,0} X \right)^{\chi/2} \otimes
\left( \Lambda^{top} T^{1,0} X \right)^{\chi/2} \: \cong \:
\left( \Lambda^{top} T^{1,0} X \right)^{2(1-g)}
\end{displaymath}
As a result, so long as $K_X^{\otimes 2} \cong {\cal O}_X$,
we see that the closed string B model is well-defined at arbitrary genus.

We shall discuss some examples of manifolds with this weaker property
in the next section, but first let us perform some checks of this claim.

First, note that Feynman-diagram-based calculations of the anomaly
only reproduce the anomaly in de Rham cohomology, and so are not
sensitive to the sorts of 2-torsion effects that are relevant here.

Next, this observation about the B model is implicit
in old standard expressions for B model correlation functions.
In for example, \cite{hubsch}, equations (C.2.9) and (C.2.12) for
B model couplings both involve not a section of the canonical
bundle $K_X$, but rather the square of such a section, or equivalently,
a section of $( K_X )^2$.  Those formulas for B model correlation functions
immediately, trivially, apply to the case when $K_X \not\cong {\cal O}_X$
but $( K_X )^{\otimes 2} \cong {\cal O}_X$.

Another bit of evidence comes from the Kodaira-Spencer theory 
\cite{bcov} describing
the string field theory of the closed string B model.  
Recall the action has the form \cite{bcov}[equ'n (5.14)]
\begin{displaymath}
\frac{1}{2} \int_X A' \frac{1}{\partial} \overline{\partial} A'
\: + \: \frac{1}{6} \int_X \left( (x + A) \wedge (x + A) \right)'
(x + A)'
\end{displaymath}
Each prime $'$ denotes contraction with a copy of the holomorphic
top-form; note that each term involves two such contractions,
{\it i.e.} the square of a section of $K_X$, or for our purposes,
a section of $K_X^{\otimes 2}$.

Before discussing some examples of manifolds with this property,
let us describe some other tests and reasons to believe that the
Calabi-Yau condition can be very slightly weakened.

Another quick test one can perform involves closure of the
closed string states under Serre duality.
Usually in string theories Serre duality has the property
of mapping mathematical descriptions of massless states
to mathematical descriptions of other massless states;
Serre duality typically defines an involution of the massless
spectrum.  In the closed string B model, it is well-known that
the massless spectrum is defined by elements of $H^{\cdot}\left(
X, \Lambda^{\cdot} T^{1,0} X \right)$.
Some simple calculations reveal
\begin{eqnarray*}
H^i\left( X, \Lambda^j T^{1,0} X \right) & \cong &
H^{n-i}\left(X, \Lambda^{n-j} \left( T^{1,0}X \right)^{\vee} \otimes
K_X \right)^* \\
& \cong &
H^{n-i}\left(X, \Lambda^j T^{1,0}X \otimes K_X^{\otimes 2} \right)^*
\end{eqnarray*}
and so we see that the massless states close back into themselves
if $K_X^{\otimes 2} = {\cal O}_X$.

As noted earlier in this paper, another motivation for this weakening
of the Calabi-Yau condition is motivated by the (0,2) generalization
of the B model.
It is straightforward to see that the B analogue twisting
of a theory describing a space $X$ with gauge bundle ${\cal E}$ is equivalent
to the A analogue twisting of a theory describing $X$ with
gauge bundle ${\cal E}^{\vee}$ -- switching ${\cal E}$ and
${\cal E}^{\vee}$ is equivalent to switching the two types
of topological twisting, as we discussed in greater detail
in section~\ref{relnAB}.  Now, as discussed in \cite{ks},
in order to make sense of the A analogue twisting of a $(0,2)$
model describing a space $X$ with gauge bundle ${\cal E}$,
we must impose the two constraints
\begin{eqnarray*}
\Lambda^{top} {\cal E}^{\vee} & \cong & K_X \\
\mbox{ch}_2( {\cal E} ) & = & \mbox{ch}_2( TX )
\end{eqnarray*}
The second of these constraints is the well-known anomaly cancellation
condition, whereas the first is a more subtle but equally important
condition.  If we take ${\cal E} = \left( T^{1,0} X \right)^{\vee}$,
which should be equivalent to working with the $(2,2)$ B model,
then we see
that the two constraints above will be satisfied if
$K_X^{\otimes 2} = {\cal O}_X$, slightly weaker than the usual
Calabi-Yau condition for the B model.

\subsection{Examples}

A well-known family of examples of complex K\"ahler manifolds with
$K_X^{\otimes 2} \cong {\cal O}_X$ but $K_X \not\cong {\cal O}_X$
are the Enriques surfaces.  These can be obtained\footnote{In fact,
it is surprisingly easy to describe the K3 covers \cite{gh}[p. 595].
Let $X$ be an
Enriques surface and $K$ the total space of the canonical bundle.
Let $s$ be a holomorphic section of $K_X^{\otimes 2}$.
Then, a K3 cover can be described as the subset of the total space
of the canonical bundle defined by
\begin{displaymath}
\left\{ (p, \kappa) | \kappa \in ( K_X)_p, \kappa^2 = s(p) \right\}
\end{displaymath}
The same is true more generally of any $n$-fold with $K^2$ trivial.}
from K3 surfaces
as quotients by freely-acting ${\bf Z}_2$'s that flip the sign of the
holomorphic top-form.  As $K_X \not\cong {\cal O}_X$, they do not
have a nowhere-zero holomorphic top-form, but since $K_X^{\otimes 2}\cong
{\cal O}_X$, in essence they have a product of holomorphic top-forms.

Enriques surfaces provide an entertaining corner case for
$(2,2)$ nonlinear sigma models.  As they are complex and K\"ahler,
a supersymmetric nonlinear sigma model on an Enriques surface has
$(2,2)$ worldsheet supersymmetry.  That is almost, but not quite,
enough to have spacetime supersymmetry.  Since Enriques surfaces
do not have nowhere-zero holomorphic top-forms, there is
no spacetime supersymmetry.  Recall \cite{banksdixon} that the condition
in a worldsheet theory for spacetime supersymmetry is a right-moving
${\cal N}=2$ algebra, plus the requirement that all physical vertex operators
have integral charge with respect to the $U(1)_R$ of the ${\cal N}=2$ algebra.
Although Enriques surfaces have $(2,2)$ worldsheet supersymmetry,
not all the physical vertex operators have integral charges.

In fact, we can see this more nearly
explicitly from the description of Enriques
surfaces as freely-acting ${\bf Z}_2$ orbifolds of K3 surfaces.
Since the ${\bf Z}_2$ acts freely, the massless modes are just the
${\bf Z}_2$-invariant massless modes of the sigma model on the K3.
The massless modes of the sigma model on the K3 are described by
the Hodge diamond of K3 cohomology:
\begin{displaymath}
\begin{array}{ccccc}
 & &  1 & & \\
 & 0 & & 0 & \\
1 & & 20 & & 1 \\
 & 0 & & 0 & \\
 & & 1 & &
\end{array}
\end{displaymath}
After performing the ${\bf Z}_2$ projection, the remaining massless
states are described by the Hodge diamond of Enriques surface cohomology:
\begin{displaymath}
\begin{array}{ccccc}
 & & 1 & & \\
 & 0 & & 0 & \\
0 & & 10 & & 0 \\
 & 0 & & 0 & \\
 & & 1 & &
\end{array}
\end{displaymath}
These still have integral charges.  However, many of the massive
states will come from the ${\bf Z}_2$-twisted sector, which because
of the twisted boundary conditions will no longer have integral charges.

This $(2,2)$ theory does not have separate left- and right-moving
spectral flow, because there is no holomorphic top-form to define
a top-charge state in chiral sectors.
Nevertheless, it does seem to have a diagonal left-right spectral flow, 
which acts
on the left- and right-movers simultaneously.
As a result, the (R,R) and (NS,NS) sectors can be mapped to one another,
but there is no map to the (R,NS) or (NS,R) sectors.
As the latter describe spacetime fermions and the former, spacetime bosons,
this seems consistent with the explicit spacetime supersymmetry breaking.

The complex structure on a Calabi-Yau can be determined in terms
of period integrals, that is, integrals of the holomorphic top-form
with respect to a basis of the middle-dimension homology.
For K3's, the resulting space of possible complex structures
can be described formally (see {\it e.g.} \cite{paulk3})
as the space of oriented 2-planes in
${\bf R}^{3,19}$ with respect to a fixed lattice $\Gamma_{3,19}$.

Although Enriques surfaces do not have a holomorphic top-form,
there is an analogous result for their moduli space of
complex structures (see {\it e.g.} \cite{allcock}).
Although the Enriques surface has no nowhere-zero holomorphic top-form with
which to define periods, since Enriques surfaces can be described
as free quotients of K3 surfaces, they inherit some of the structure
of K3 surfaces.
At least locally complex structures can be described in terms
of oriented 2-planes in ${\bf R}^{2,10}$, with respect to a fixed
lattice $\Gamma_{2,10}$.

In passing, hyperelliptic (sometimes called bielliptic) surfaces
also have torsion canonical bundle, and in some cases $( K_X )^{\otimes 2}
= {\cal O}_X$.  These surfaces are obtained as freely-acting
orbifolds of products of
elliptic curves.  See \cite{beauville}[list VI.20],
\cite{gh}[pp. 585-590] for more information.
These surfaces are distinct from Enriques surfaces, though the description
might incorrectly make them sound like orbifold limits thereof.
Because they are freely-acting orbifolds of Calabi-Yau's, the analysis
of the worldsheet physics in these cases is very similar to that for
the Enriques surfaces just described.

Another easy example of a complex K\"ahler manifold with
the property that $(K_X)^{\otimes 2} \cong {\cal O}_X$
but $K_X \not\cong {\cal O}_X$ can be obtained as follows.
Let $E$ be an elliptic curve, and let $L$ be a flat line bundle over
$E$ whose holomorphic structure is determined by a 2-torsion point of
$E$.  Then, the total space of $L$ has the property that
$(K_X)^{\otimes 2} \cong {\cal O}_X$ but $K_X \not\cong {\cal O}_X$.
This space has a double cover which is Calabi-Yau, namely
$E \times {\bf C}$.  The space with $K \not\cong {\cal O}$ is
obtained by quotienting by a freely-acting ${\bf Z}_2$ that
acts as translation by a 2-torsion point on $E$ combined with
$x \mapsto -x$ on ${\bf C}$.

More generally, if a complex K\"ahler manifold $X$ has the property
that $K_X^{\otimes 2} \cong {\cal O}_X$ but $K_X \not\cong {\cal O}_X$,
then \cite{donagipriv} $X$
is not simply-connected, and it has a double cover which is
a Calabi-Yau.  For this reason, ultimately these results on extending
the ordinary B model to more general manifolds do not strike the authors
as being overly important -- all examples descend from Calabi-Yau's.
However, because the general condition crops up in our analysis of
the (0,2) B model, we do feel it is important to describe it here.

\subsection{Mirror symmetry}

The mirrors to Enriques surfaces and other complex manifolds
with $K^2$ trivial can be described very easily.
Recall from the last section that every complex K\"ahler manifold
with $K^2$ trivial but $K$ nontrivial has a double cover which is
a Calabi-Yau.  

Now, if a Calabi-Yau admits a holomorphic ${\bf Z}_2$ involution which
flips the sign of the holomorphic top-form, then its mirror
necessarily must have the same property.
Physically this is a consequence of the conformal field theory:
existence of such a ${\bf Z}_2$ is a property of the CFT,
and since mirrors have the same CFT, if one admits such a ${\bf Z}_2$,
then the other must also.
Mathematically, this corresponds to a known fact concerning K3's,
responsible for certain properties of Voisin-Borcea manifolds.
More generally, this can be understood mathematically if the
Calabi-Yau is described with a special Lagrangian torus fibration, as follows.
(We would like to thank M.~Gross for providing this argument.)
The monodromy of the fibration, in general, is represented in $GL(n,{\bf Z})$,
and the canonical class is trivial if and only if the monodromy lies
in $SL(n,{\bf Z})$.  Mirror symmetry in this language dualizes the
monodromy representation, so the monodromy representation of the mirror
will lie in $SL(n,{\bf Z})$ if and only if the monodromy representation of
the original lays in $SL(n,{\bf Z})$.  If the canonical class is 2-torsion,
then it does not lie in $SL(n,{\bf Z})$, and cannot be made to lie in
$SL(n,{\bf Z})$ after duality, though the dual will have the property that
it is again 2-torsion.

The result should now be clear.
The mirror to an Enriques surface or other $n$-fold with
$K^2$ trivial, is another Enriques surface or $n$-fold with $K^2$
trivial, constructed by lifting to a Calabi-Yau cover, taking the
mirror of that Calabi-Yau cover, and then quotienting. 
After all, since existence of the ${\bf Z}_2$ involution on the
Calabi-Yau cover can be understood as a property of the CFT
of the Calabi-Yau, and the CFT is invariant under mirror
symmetry, the mirror to the ${\bf Z}_2$ orbifold 
must be the orbifold of the mirror Calabi-Yau.

\subsection{Open string B model}

The weakening of the Calabi-Yau condition we have just outlined
is specific to the {\it closed} string B model.
Consider the open string B model describing, for example,
open strings connecting
a B-brane wrapped on all of the complex manifold $X$i back to itself.
The boundary conditions on the fermions kill half of the
fermion zero modes, so to make the path integral measure well-defined,
we need a holomorphic section of only one factor of
$\Lambda^{top} T^{1,0} X$, which implies the usual Calabi-Yau condition.

Since open strings encode closed strings, if we had gotten
a weaker condition for open strings than closed strings, we would
be in trouble; but since the condition for open strings is stronger
than the condition for closed strings, we seem to be consistent.

\section{Anomaly analysis}
\label{anomanalysis}

\subsection{Genus zero}    \label{anomg0}

In appendix~\ref{bmod} we analyzed $(2,2)$ A and B model anomalies classically.
Let us repeat that calculation for the $(0,2)$ generalization of the
A model first discussed in \cite{ks}, and then extend that calculation
to quantum sectors, to verify that the path integral measure is always
well-defined. 

Classically recall if the target space is $X$, and the gauge bundle
is ${\cal E}$, then the worldsheet fermions in the A-type twist
of a $(0,2)$ model are defined by the following bundles:
\begin{eqnarray*}
\psi_+^i & \in & \Gamma_{ C^{\infty} }\left( \phi^* T^{1,0}X \right) \\
\psi_+^{\overline{\imath}} & \in &
\Gamma_{ C^{\infty} }\left( K_{\Sigma} \otimes \left(
\phi^* T^{1,0}X \right)^{\vee} \right) \\
\lambda_-^a & \in & \Gamma_{ C^{\infty} }\left(
\overline{K}_{\Sigma} \otimes \left( \phi^* \overline{\epsilon} \right)^{\vee} 
\right) \\
\lambda_-^{\overline{a}} & \in &
\Gamma_{ C^{\infty} }\left( \phi^* \overline{\epsilon} \right)
\end{eqnarray*}
The zero modes of the $\psi_+$ are holomorphic sections of the
indicated bundles, and the zero modes of the $\lambda_-$ are antiholomorphic
sections of the indicated bundles.
In a classical sector, $\phi^* {\cal G}$ for any holomorphic bundle
${\cal G}$ is a trivial bundle over the worldsheet whose rank matches
that of ${\cal G}$, so if the worldsheet is ${\bf P}^1$, then
we see only the $\psi_+^i$ and $\lambda_-^{\overline{a}}$ have zero modes.
To make the path-integral measure well-defined, we would appear
to need a holomorphic section of $\Lambda^{top} T^{1,0}X$ and an antiholomorphic
section of $\Lambda^{top} \overline{ {\cal E} }$.
However, in the definition of the sigma model action there is also
implicitly a choice of hermitian fiber metric on the vector bundle 
${\cal E}$, and just as in appendix~\ref{bmod}, we can use that metric to 
dualize
the antiholomorphic vector bundle $\Lambda^{top} \overline{ {\cal E} }$
into the holomorphic vector bundle $\Lambda^{top} {\cal E}^{\vee}$,
as smooth bundles.
Just as in appendix~\ref{bmod}, in order for the path integral measure to be
well-defined, all we really need is a nowhere-zero holomorphic
section of the product
of these two bundles
\begin{displaymath}
\Lambda^{top} {\cal E}^{\vee} \otimes \Lambda^{top} T^{1,0} X
\end{displaymath}
One of the constraints on these theories is that
$\Lambda^{top} {\cal E}^{\vee} \cong K_X$, which implies that the product
above is the trivial bundle ${\cal O}_X$, which always has a section,
and so we get no new constraints.

Now let us repeat this analysis for quantum sectors.
In a sector where the space of bosonic zero modes is 
${\cal M}$, in the A-type twist of a $(0,2)$ theory the
worldsheet fermions are interpreted in terms of the following
bundles:
\begin{eqnarray*}
\psi_+^i & \in & \Gamma_{ C^{\infty} }\left( R^0 \pi_* \alpha^* 
T^{1,0} X \right) \: \cong \: \Gamma_{ C^{\infty} }\left(
T {\cal M} \right) \\
\psi_+^{\overline{\imath}} & \in & \Gamma_{ C^{\infty} }\left(
\left( R^1 \pi^* \alpha^* T^{1,0} X \right)^{\vee} \right) \: \cong \:
\Gamma_{ C^{\infty} }\left( \left( \mbox{Obs} \right)^{\vee}\right) \\
\lambda_-^a & \in & \Gamma_{ C^{\infty} }\left(
\left( \overline{ R^1 \pi_* \alpha^* {\cal E} } \right)^{\vee} 
\right) \: \cong \:
\Gamma_{ C^{\infty} }\left( \left( \overline{ {\cal F}_1 } \right)^{\vee}
\right) \\
\lambda_-^{\overline{a}} & \in & \Gamma_{ C^{\infty} }\left(
\overline{ R^0 \pi_* \alpha^* {\cal E} } \right) \: \cong \:
\Gamma_{ C^{\infty} }\left( \overline{ {\cal F} } \right)
\end{eqnarray*}
and, as before, the zero modes of the $\psi_+$ are holomorphic sections
of the indicated bundles, whereas the zero modes of the $\lambda_-$ are
antiholomorphic sections of the indicated bundles.

As before, there is a hermitian fiber metric which we can use to
dualize the antiholomorphic vector bundle
$\overline{ {\cal F} }$ into the holomorphic vector bundle ${\cal F}^{\vee}$,
and also $\left( \overline{ {\cal F}_1 } \right)^{\vee}$ into 
${\cal F}_1$, as smooth bundles.
As before, to make the path integral measure well-defined, we must 
require that there exist a holomorphic nowhere-zero section of
the product
\begin{displaymath}
\left( \Lambda^{top} T {\cal M} \right) \otimes
\left( \Lambda^{top} \left( \mbox{Obs} \right)^{\vee} \right) \otimes
\left( \Lambda^{top} {\cal F}_1 \right) \otimes
\left( \Lambda^{top} {\cal F}^{\vee} \right)
\end{displaymath}
However, in \cite{ks} we saw that the constraints
\begin{eqnarray*}
\Lambda^{top} {\cal E}^{\vee} & \cong & K_X \\
\mbox{ch}_2({\cal E}) & = & \mbox{ch}_2(TX) 
\end{eqnarray*}
together with Grothendieck-Riemann-Roch guarantee\footnote{Technically,
GRR only shows that the result is true topologically, but in all examples
considered to date the result is true holomorphically.}
that the product of line bundles above is isomorphic to ${\cal O}_{ {\cal M} }$,
which always has a section.
(Strictly speaking, one has to compactify ${\cal M}$ and extend the
sheaves above over the compactification, in a fashion preserving symmetries,
as we discussed in \cite{ks}, and there we discussed how this could be done,
in such a way as to insure that the product of four line bundles above
is isomorphic to the trivial line bundle even over the compactification.)

Thus, we find that in our $(0,2)$ analogue of the A model
on worldsheet ${\bf P}^1$, so long
as the constraints 
\begin{eqnarray*}
\Lambda^{top} {\cal E}^{\vee} & \cong & K_X \\
\mbox{ch}_2({\cal E}) & = & \mbox{ch}_2(TX)
\end{eqnarray*}
are obeyed, the path integral measure is well-defined, even in
quantum sectors.

\subsection{Higher genera}    \label{anomhg}

The same analysis can also be repeated for higher genus
Riemann surfaces.  We will {\it not} couple to topological gravity
here, as it introduces complications\footnote{For example,
strictly speaking correlation functions contain a ratio of operator
determinants.  For fixed complex structure, this is just a number,
which we can ignore, but if we couple to topological gravity,
then we must take into account its functional dependence.} 
beyond the scope of the present paper.

First, consider the classical case at arbitrary genus.
To make the path-integral measure well-defined, we need a holomorphic 
nowhere-zero section of 
\begin{displaymath}
\left( \Lambda^{top} T^{1,0} X\right)^{\chi/2} \otimes
\left( \Lambda^{top} {\cal E} \right)^{-\chi/2}
\end{displaymath}
(The $\psi_+^i$ zero modes transform as $h^0\left(\Sigma, {\cal O}\right)$
copies of $T^{1,0} X$, whereas the $\psi_+^{\overline{\imath}}$ zero
modes transform as $h^1\left(\Sigma, {\cal O}\right)$ copies of 
$\left( T^{1,0} X \right)^{\vee}$, so they require the
factor of 
\begin{displaymath}
h^0\left( \Sigma, {\cal O} \right) \: - \:
h^1\left( \Sigma, {\cal O} \right) \: = \: \chi({\cal O}) \: = \:
\chi/2
\end{displaymath}
powers of $\Lambda^{top} T^{1,0} X$, and similarly for the left-movers.)
This condition generalizes the constraint $\Lambda^{top} {\cal E}^{\vee}
\: \cong \: K_X$ that appeared at genus zero.
Also note that when ${\cal E} \cong \left( T^{1,0} X \right)^{\vee}$,
{\it i.e.} the (2,2) B model locus, the condition above becomes
that $K_X^{\otimes \chi} \cong {\cal O}_X$, which is the correct
higher-genus consistency condition for the closed string B model,
as discussed in appendix~\ref{bmod}.
We also, of course, need the standard anomaly cancellation condition
\begin{displaymath}
\mbox{ch}_2({\cal E}) \: = \:
\mbox{ch}_2({\cal TX})
\end{displaymath}

Quantum mechanically, in order for the path integral measure to
be well-defined we require a holomorphic nowhere-zero section of
\begin{displaymath}
\left( \Lambda^{top} R^0 \pi_* \alpha^* T^{1,0} X \right) \otimes
\left( \Lambda^{top} R^1 \pi_* \alpha^* T^{1,0} X \right)^{\vee} \otimes
\left( \Lambda^{top} R^0 \pi_* \alpha^* {\cal E} \right)^{\vee} \otimes
\left( \Lambda^{top} R^1 \pi_* \alpha^* {\cal E} \right)
\end{displaymath}
(which reduces to the classical condition in the zero instanton sector).
Just as in \cite{ks}, this constraint\footnote{Strictly speaking,
Grothendieck-Riemann-Roch reproduces this constraint topologically.
However, in all examples discussed in \cite{ks} and here,
this constraint held true holomorphically, and furthermore 
compactifications and extensions of the sheaves could be chosen so as
to preserve this condition holomorphically as well.}
is a consequence of the classical conditions and Grothendieck-Riemann-Roch.
The analysis is very similar to that in \cite{ks}; we briefly review
it here for completeness.
Letting a subscript of $k$ on a cohomology
class denote its complex codimension $k$ component, and letting $\eta$
be the pullback to $\Sigma\times {\cal M}$ of the cohomology class of a
point of $\Sigma$, we have
\begin{eqnarray}  
c_1\left(R^0 \pi_* \alpha^* {\cal E} \ominus
R^1 \pi_* \alpha^* {\cal E}\right) 
& = &
\pi_*\left( \left( \mbox{ch}(\alpha^* {\cal E}) \mbox{Td}(T \Sigma) \right)_2
\right) \nonumber\\
& = & \pi_*\left( \alpha^* \left( \mbox{ch}_2({\cal E})
 \right) + \left(1-g\right)\eta\alpha^*c_1({\cal E})\right).
\label{grr1}
\end{eqnarray}
If we apply the result above to ${\cal E} = TX$ we obtain
\begin{eqnarray*}
c_1\left( R^0 \pi_* \alpha^* T^{1,0} X \ominus R^1 \pi_* \alpha^*
T^{1,0}X \right) 
& = &
\pi_*\left( \left( \mbox{ch}(\alpha^* TX) \mbox{Td}(T \Sigma) \right)_2
\right) \\
& = & \pi_*\left( \alpha^* \left( \mbox{ch}_2(TX)
 \right) + \left(1-g\right)\eta\alpha^*c_1(TX)\right).
\end{eqnarray*}
The condition that there be a nowhere-zero holomorphic section of
\begin{displaymath}
\left( \Lambda^{top} T^{1,0} X\right)^{\chi/2} \otimes
\left( \Lambda^{top} {\cal E} \right)^{-\chi/2}
\end{displaymath}
implies that
\begin{displaymath}
(1 - g) c_1({\cal E}) \: = \: (1 - g) c_1( T^{1,0} X )
\end{displaymath}
which together with the anomaly cancellation condition and the results
above tells us that
\begin{displaymath}
c_1\left( R^0 \pi_* \alpha^* {\cal E} \ominus
R^1 \pi_* \alpha^* {\cal E}\right) \: = \:
c_1\left( R^0 \pi_* \alpha^* T^{1,0} X \ominus R^1 \pi_* \alpha^*
T^{1,0}X \right) 
\end{displaymath}
which is the (topological) version of the constraint that
there be a nowhere-zero holomorphic section of 
\begin{displaymath}
\left( \Lambda^{top} R^0 \pi_* \alpha^* T^{1,0} X \right) \otimes
\left( \Lambda^{top} R^1 \pi_* \alpha^* T^{1,0} X \right)^{\vee} \otimes
\left( \Lambda^{top} R^0 \pi_* \alpha^* {\cal E} \right)^{\vee} \otimes
\left( \Lambda^{top} R^1 \pi_* \alpha^* {\cal E} \right)
\end{displaymath}
Thus, the condition we need for the path integral measure to be
well-defined in nonzero instanton sectors at higher genus, is a consequence
of the conditions at zero instanton number, just as happened at genus zero.

\end{document}